\documentclass[aps,prc,preprint,superscriptaddress,showkeys]{revtex4-2}
\usepackage[T1]{fontenc}
\usepackage{graphicx}
\usepackage{dcolumn}
\usepackage{bm}
\usepackage{amsmath}

\begin{document}

\title{Many-channel microscopic cluster model of $^{8}$Be: S-factors}

\author{V. I. Zhaba}
\email{viktorzh@meta.ua}
\affiliation{Bogolyubov Institute for Theoretical Physics, 03143 Kyiv, Ukraine}
\author{Yu. A. Lashko}
\email{ylashko@gmail.com}
\affiliation{Bogolyubov Institute for Theoretical Physics, 03143 Kyiv, Ukraine}
\affiliation{National Institute for Nuclear Physics, Padova Division, 35131 Padova, Italy}
\author{V. S. Vasilevsky}
\email{vsvasilevsky@gmail.com}
\affiliation{Bogolyubov Institute for Theoretical Physics, 03143 Kyiv, Ukraine}

\date{\today}

\begin{abstract}
We investigate low--energy astrophysical $S$ factors for reactions proceeding through the $^{8}$Be compound system with entrance channels $p+{}^{7}$Li, $n+{}^{7}$Be, and $d+{}^{6}$Li. Using the same microscopic many--channel three--cluster framework as in our previous study of the high--lying $^{8}$Be spectrum, we calculate $S(E)$ for $^{7}$Li($p,\alpha)^{4}$He, $^{7}$Be($n,\alpha)^{4}$He, $^{7}$Be($n,p)^{7}$Li, $^{6}$Li($d,\alpha)^{4}$He,
$^{6}$Li($d,p)^{7}$Li, and $^{6}$Li($d,n)^{7}$Be in the energy range relevant for primordial and stellar nucleosynthesis. For the mirror pair $^{7}$Li($p,\alpha)^{4}$He / $^{7}$Be($n,\alpha)^{4}$He and for $^{7}$Be($n,p)^{7}$Li the calculated $S$ factors reproduce both the absolute scale and the low--energy trends of the experimental data within their quoted uncertainties, whereas the absolute $S$ factors for the deuteron--induced channels on $^{6}$Li are underestimated at low energy, consistent with the
shifted $^{6}$Li+$d$ threshold and the absence of a broad subthreshold $2^{+}$ structure in the present implementation. A partial--wave analysis identifies the dominant $J^{\pi}$ contributions in each channel and relates them to specific $^{8}$Be resonances, while demonstrating that cluster polarization, previously shown to be crucial for the $^{8}$Be spectrum, is likewise
essential for the normalization and energy dependence of several $S$ factors. Evaluating $S(E)$ at appropriate Gamow energies, we obtain a hierarchy of reaction channels that quantifies the relative importance of neutron-- and deuteron--induced processes for the production and destruction of $^{7}$Li and $^{7}$Be.
\end{abstract}

\maketitle

\section{Introduction}
\label{intro}

Reactions that proceed through the compound nucleus $^{8}$Be at low
entrance--channel energies (tens to a few hundred keV) are key inputs for
astrophysical reaction rates. They also provide a stringent test of cluster
models, because multiple binary partitions couple through $^{8}$Be and there
exists a broad set of low--energy cross--section and $S$--factor data, while
a unified microscopic description of these observables is still lacking.
Among these, reactions involving $^{7}$Li and $^{7}$Be enter standard
calculations of light--element nucleosynthesis and are directly relevant to
the long--standing cosmological lithium problem.

In our previous work~\cite{2025PhRvC.112a4328Z} (hereafter Paper~I), we
analyzed high--excitation resonances of $^{8}$Be within a microscopic
three--cluster model including the configurations
\[
^{4}\text{He}+^{3}\text{H}+p,\quad
^{4}\text{He}+^{3}\text{He}+n,\quad
^{4}\text{He}+d+d,\quad
^{4}\text{He}+2p+2n,
\]
which embed all binary rearrangement channels
\[
^{4}{\rm He}+^{4}{\rm He},\;
^{7}{\rm Li}+p,\;
^3{\rm H}+^5{\rm Li},\;
^{7}{\rm Be}+n,\;
^3{\rm He}+^5{\rm He},\;
d+^{6}{\rm Li},\;
2n+^6{\rm Be},\;
2p+^6\rm{He}.
\]
That study established the rich spectrum of $^{8}$Be near the
$p+{}^{7}$Li, $n+{}^{7}$Be, and $d+{}^{6}$Li thresholds, and demonstrated that
cluster polarization in the $A=5$, 6, and 7 subsystems is decisive for the
formation and properties of the twin $1^{+}$, $2^{+}$, $3^{+}$, and $4^{+}$
resonance doublets.

Here we apply the same microscopic many--channel three--cluster framework to
reactions at low entrance--channel energies in the
$p+{}^{7}$Li, $n+{}^{7}$Be, and $d+{}^{6}$Li channels, and focus on the
astrophysical $S$ factors for
\begin{eqnarray}
^{7}\text{Li}\left(  p,\alpha\right)  ^{4}\text{He}, \quad
^{7}\text{Be}\left( n,\alpha\right)  ^{4}\text{He},\quad
^{7}\text{Be}\left(  n,p\right)  ^{7}\text{Li}, \label{eq:000} \\
^{6}\text{Li}\left(  d,\alpha\right)  ^{4}\text{He},\quad
^{6}\text{Li}\left(  d,p\right)  ^{7}\text{Li}, \quad
^{6}\text{Li}\left(  d,n\right) ^{7}\text{Be}. \nonumber
\end{eqnarray}
Our goal is a unified microscopic description of the six reactions in
Eq.~(\ref{eq:000}) that (i) predicts low--energy $S(E)$ from explicit cluster
structure and intercluster dynamics, (ii) identifies the dominant $J^{\pi}$
contributions to each $S$ factor and links them to specific $^{8}$Be
resonances found in Paper~I, and (iii) quantifies how cluster polarization
controls the absolute scale and energy dependence of $S(E)$ in the
astrophysical regime.

The $^{8}$Be spectrum is rich: many states lie above the $p+^{7}$Li threshold
($E_x\gtrsim17$~MeV), while the $0^{+}$ ground state is a very narrow
resonance just above the $^{4}\mathrm{He}+^{4}\mathrm{He}$ threshold. Because
several open and near--threshold partitions are coupled through $^{8}$Be,
the reactions in Eq.~(\ref{eq:000}) have attracted sustained experimental
effort. Cross sections and $S$ factors have been measured with improving
precision down to tens of keV, and low--energy extrapolations commonly use
linear or low--order polynomial parametrizations of $S(E)$. A reasonably
consistent picture emerges across different experiments and techniques.
Appendix~\ref{Sec:Exper} summarizes the experimental datasets, within the
energy range considered in this work, that are used in our $S$-factor
comparisons for the six reactions listed above.

By contrast, modern microscopic calculations of these $S$ factors remain
comparatively sparse and reaction--specific. Descouvemont and Baye treated
$^{7}$Li$(p,\alpha)^{4}$He, $^{7}$Be$(p,n)^{7}$Li, and $^{7}$Li$(p,\gamma)^{8}$Be in a generator--coordinate three--cluster model
\cite{1994NuPhA.573...28D}. While total cross sections were described
over a broad energy range, only the $S$ factor for
$p+^{7}$Li$\to\alpha+\alpha$ was analyzed in the astrophysical limit; the
calculation overestimates both the absolute scale and the low--energy slope,
predicting a pronounced rise of $S(E)$ as $E\!\to\!0$ driven by the twin
$2^{+}$ subthreshold states of $^{8}$Be just below the $p+^{7}$Li threshold,
whereas the experimental $S(E)$ is almost flat with at most a weak upward
trend.

A microscopic many--cluster calculation for $^{7}$Li$(p,\gamma)^{8}$Be yields
$S(E)$ for radiative capture \cite{1996NuPhA.607...62C}, while DWBA analyses
have been performed for $^{7}$Li$(p,\alpha)^{4}$He
\cite{1990PhLB..249..191R,1995NuPhA.582..270Y} and
$^{6}$Li$(d,\alpha)^{4}$He~\cite{Ruprecht2004}. In particular,
Ref.~\cite{1995NuPhA.582..270Y} showed that finite--range DWBA with
direct+exchange amplitudes can reproduce low--energy angular distributions and
provide an $S(E)$ curve consistent with existing data, and
Ref.~\cite{Ruprecht2004} emphasized the importance of coherently combining
direct and resonant amplitudes near threshold, driven by interference with a
subthreshold $2^{+}$ resonance in $^{8}$Be. These studies reinforce the
conclusion that low--energy observables in these systems are highly sensitive
to near--threshold structure and benefit from models with explicit cluster
dynamics.

Complementary to these structure--based approaches, global $R$-matrix
evaluations of the $^{8}$Be system provide a phenomenological, data--driven
description. An early multichannel analysis by Page and collaborators fitted a
broad set of scattering and reaction data and yielded differential and
integrated cross sections for many of the channels listed in
Eq.~(\ref{eq:000})~\cite{2005PhRvC..72e4312P,Page20052}. More recently,
Paneru \textit{et al.} carried out a comprehensive Bayesian $R$-matrix
analysis using the AZURE2 code~\cite{2025PhRvC.111f4609P}, assimilating new
datasets (in particular for deuteron--induced reactions on $^{6}$Li),
quantifying uncertainties for level and channel parameters, and providing
evaluated cross sections for several of the reactions in
Eq.~(\ref{eq:000}). As a phenomenological framework, the $R$-matrix approach
is well suited to interpolation, extrapolation, and reaction--rate evaluations;
however, it does not by itself resolve the underlying cluster dynamics or the
role of subthreshold structure across all entrance and exit channels.

Thus, the gap we address here is a unified, microscopic and multichannel
description of the six reactions in Eq.~(\ref{eq:000}) that simultaneously
connects to the high--lying $^{8}$Be resonances identified in Paper~I and
predicts low--energy $S(E)$ in the astrophysical entrance--channel regime.

The paper is organized as follows. In Sec.~\ref{Model} we summarize the many--channel
microscopic three--cluster model for $^{8}$Be and its implementation for the
relevant binary partitions, including the definition of the astrophysically
relevant Gamow windows for the reactions of interest. Section~\ref{S factors} presents the
calculated astrophysical $S$ factors and compares them with available
experimental data, highlighting the role of Coulomb effects in mirror
channels, the low--energy behavior and analytical approximations to $S(E)$,
and the resulting reaction hierarchy at representative Gamow energies. Our
main findings and their astrophysical implications are summarized in Sec.~\ref{concl}, 
and Appendix~\ref{Sec:Exper} lists the experimental datasets used in
our $S$-factor comparisons.

\section{Many-channel microscopic model \label{Model}}
\subsection{Cluster configurations and nucleon--nucleon interaction}

To study nuclear reactions proceeding through the compound nucleus $^{8}$Be, we employ the same microscopic three–cluster model as in Ref.~ \cite{2025PhRvC.112a4328Z}, previously used to describe high–energy resonance states of $^{8}$Be. This model explicitly includes all eight binary partitions of $^{8}$Be:
\[
^{4}\text{He}+^{4}\text{He},\quad p+^{7}\text{Li},\quad n+^{7}\text{Be}%
,\quad d+^{6}\text{Li},\text{ }2n\text{+}^{6}\text{Be},\quad 2p\text{+}%
^{6}\text{He},\quad t\text{+}^{5}\text{Li},\quad ^{3}\text{He+}^{5}\text{He.}
\]
Consequently, the total wave function of a continuum state of $^{8}$Be is written as a superposition of eight channel wave functions, each describing one of these binary partitions.

The wave function of $^{8}$Be in a given three–cluster configuration can be represented as
\begin{equation}
\Psi ^{\left( E,J\right) }=\sum_{\alpha }\widehat{\mathcal{A}}\left\{ \Phi
_{\alpha }\left( A_{\alpha },S_{\alpha }\right) \psi _{\alpha }\left( 
\mathcal{E}_{\alpha },A_{\beta }+A_{\gamma },j_{2}\right) \varphi _{E-%
\mathcal{E}_{\alpha },l_{1},j_{1}}\left( \mathbf{y}_{\alpha }\right)
\right\} _{J},
\label{eq:M001}
\end{equation}
where $\Psi^{(E,J)}$ denotes the wave function of $^{8}$Be with total energy $E$
and total angular momentum $J$. The index $\alpha$ labels the three clusters in a given
three–cluster configuration, with $(\alpha,\beta,\gamma)$ running over all permutations of
the three constituents. Thus, for a fixed three–cluster configuration the sum over $\alpha$
represents the three binary channels
$A_{\alpha} + (A_{\beta}+A_{\gamma})$, each term describing the scattering of the
cluster with index $\alpha$ on a bound or pseudo–bound two–cluster state formed by the
clusters $\beta$ and $\gamma$.

The function $\Phi_{\alpha}(A_{\alpha},S_{\alpha})$ is the many–particle shell–model wave
function that describes the internal structure of the “projectile’’ cluster $A_{\alpha}$,
with intrinsic spin $S_{\alpha}$. In the present model the projectiles are the proton,
neutron, deuteron, $2n$, $2p$, triton, $^{3}$He, and the $^{4}$He cluster. They are treated
microscopically as composites of nucleons but, in contrast to the target nuclei, are not
themselves described as two–cluster subsystems. The function
$\psi_{\alpha}(\mathcal{E}_{\alpha},A_{\beta}+A_{\gamma},j_{2})$ describes the internal
structure of the “target’’ nucleus, i.e., a two–cluster system
($\alpha+n$, $\alpha+p$, $\alpha+d$, $\alpha+2n$, $\alpha+2p$, $\alpha+t$, $\alpha+{}^{3}$He)
corresponding to $^{5}$He, $^{5}$Li, $^{6}$Li, $^{6}$He, $^{6}$Be, $^{7}$Li, or $^{7}$Be.
Here $\mathcal{E}_{\alpha}$ is the internal energy of this
two–cluster subsystem, and $j_{2}$ is its total angular momentum.

The relative–motion function
$\varphi_{E-\mathcal{E}_{\alpha },l_{1},j_{1}}(\mathbf{y}_{\alpha})$ describes the motion of
the projectile $A_{\alpha}$ with respect to the center of mass of the pair
$A_{\beta}+A_{\gamma}$. The vector $\mathbf{y}_{\alpha}$ is the corresponding Jacobi
coordinate, $l_{1}$ is the orbital angular momentum associated with this relative motion,
and $j_{1}$ is the total angular momentum obtained by coupling $l_{1}$ to the channel
spin. The functions $\psi_{\alpha}$ and $\Psi^{(E,J)}$ are obtained as solutions of the
corresponding Schrödinger equations with appropriate boundary conditions. The elements of
the scattering $S$ matrix can be extracted from the asymptotic behavior of
$\varphi_{E-\mathcal{E}_{\alpha },l_{1},j_{1}}(\mathbf{y}_{\alpha})$ at large intercluster
separations ($|\mathbf{y}_{\alpha}|\gg 1$); an equivalent, more rigorous procedure is to
solve the system of inhomogeneous linear equations given in Eq.~(16) of
Ref.~\cite{2024PhRvC.109d5803L}. Further details of the microscopic many–channel model
can be found in Refs.~\cite{2024PhRvC.109d5803L,2025PhRvC.112a4328Z}.

The astrophysical $S$ factors for reactions in the compound nucleus $^{8}$Be are calculated using the Hasegawa–Nagata nucleon–nucleon potential \cite{potMHN1,potMHN2}. We adopt the same potential and input parameters as in Paper~I \cite{2025PhRvC.112a4328Z}, where the Majorana parameter was slightly adjusted to reproduce the energies of the ground ($3/2^{-}$) and first excited ($1/2^{-}$) states of $^{7}$Li and $^{7}$Be, and thereby the correct relative positions of the $p+{}^{7}$Li and $n+{}^{7}$Be thresholds. Because these nuclei and their low–lying states play a central role in reactions of astrophysical interest, particular care was taken to describe their structure accurately. With this choice, however, the $d$ + $^{6}$Li threshold cannot be reproduced simultaneously with the $p+{}^{7}$Li and $n+{}^{7}$Be thresholds; the implications of this limitation for deuteron–induced reactions on $^{6}$Li will be discussed below.

\subsection{Astrophysical Gamow energies and reaction windows}

The energy range most relevant for thermonuclear reactions is defined by the Gamow window, which identifies the region where the product of the Maxwell-Boltzmann distribution and the tunneling probability is maximized \cite{1999NuPhA.656....3A}. This concept provides a practical criterion for estimating the energies at which charged-particle-induced reactions predominantly occur in astrophysical environments.

In Table~\ref{Tab:GamowWinds}, we present the Gamow peak energy \( E_0 \) and the corresponding width \( \Delta E_0 \) for all entrance channels associated with the reactions considered in this work.
Following the approach adopted in our previous studies \cite{2024PhRvC.109d5803L,2024PhRvC.110c5806L}, these parameters are evaluated for a typical astrophysical temperature of \( T_9 = 0.8 \), where \( T_9 \) denotes the temperature in units of \( 10^9 \) K.
 
For reactions involving charged particles or clusters, we employ the standard analytic expressions for the Gamow peak and width:
\begin{eqnarray}
E_{0}& =& 0.122\left( {Z_{1}^{2} Z_{2}^{2} \frac{A_{1} A_{2} }{A_{1} +A_{2}
}T_{9}^{2} } \right)^{1/3} {\rm MeV}, \label{eq:M101A}\\
\Delta E_{0} &=&0.2368 \left( {Z_{1}^{2} Z_{2}^{2} \frac{A_{1} A_{2} } {A_{1} +A_{2}
 }T_{9}^{5} } \right)^{1/6} {\rm MeV}, \label{eq:M101B}
\end{eqnarray}
where $A_{1}$ and $Z_{1}$ are the mass number and charge of the first interacting nucleus, and $A_{2}$ and $Z_{2}$ are those of the second.

For reactions induced by neutron–nucleus interactions, an analog of the Gamow window can be used, referred to as the effective-energy window. As defined in Ref.~\cite{1969ApJS...18..247W}, the effective-energy window is given by $E_0 \pm \frac{1}{2} \Delta E_0$, where
\begin{eqnarray}
E_{0}  & =0&.086\left(  l+\frac{1}{2}\right)  T_{9},\label{eq:M102A}\\
\Delta E_{0}  & =& 0.097\left(  l+\frac{1}{2}\right)  ^{1/2}T_{9}%
,\label{eq:M102B}%
\end{eqnarray}
and $l$ is the orbital angular momentum of the relative motion between the interacting nuclei.

The cross sections of the $^{7}$Be$(n,\alpha)^4$He reaction for $p$-wave neutrons were experimentally determined for the first time in Ref.~\cite{2017PhRvL.118e2701K}, at $E_{\text{cm}} = 0.20$–$0.81$ MeV, slightly above the effective-energy window, by applying the principle of detailed balance to the time-reverse reaction.

In all subsequent figures, the Gamow (effective–energy) window is shown as a dashed band in the astrophysical $S$-factor plots for each reaction. For the calculation of the effective–energy window parameters $E_{0}$ and $\Delta E_{0}$ for the reaction induced by the $n+{}^{7}$Be interaction, the orbital angular momentum was taken as $l=1$, corresponding to the dominant partial wave in the entrance channel at low energies.

\begin{table}[tbp] \centering
\caption{Gamow peak energy $E_{0}$ and width $\Delta E_{0}$ for the reaction entrance channels, calculated at $ T = 0.8$ GK.}%
\begin{tabular}
[c]{|c|c|c|}\hline
Channel & $E_{0}$, keV & $\Delta E_{0}$, keV\\\hline
$p$+$^{7}$Li & 209 & 277\\\hline
$d$+$^{6}$Li & 250 & 303\\\hline
$^{3}$He+$^{5}$He & 327 & 347\\\hline
$n$+$^{7}$Be & 103 & 95\\\hline
\end{tabular}
\label{Tab:GamowWinds}%
\end{table}%

\section{Astrophysical $S$ factors and comparison with experiment}
\label{S factors}

In this section we present the astrophysical $S$ factors obtained with the
many--channel microscopic model of Sec.~\ref{Model} and compare them with available
experimental data. We first discuss reactions on $A=7$ nuclei, beginning with
the $\alpha$--emission mirror pair $^{7}$Li$(p,\alpha)^{4}$He and
$^{7}$Be$(n,\alpha)^{4}$He, which share the same $\alpha+\alpha$ exit channel
and therefore involve only even-$J$, positive--parity states, and then turn to
the charge--exchange reaction $^{7}$Be$(n,p)^{7}$Li. We then consider the
deuteron--induced reactions on $^{6}$Li. Global systematics, including Coulomb
effects in the charged entrance channels, low--energy approximations to
$S(E)$, and the reaction hierarchy at the Gamow energies defined in
Sec.~\ref{Model}, are discussed in the subsequent subsections.

\subsection{$\alpha$--emission reactions: $^{7}$Li$(p,\alpha)^{4}$He and $^{7}$Be$(n,\alpha)^{4}$He}

Because the exit channel is $\alpha+\alpha$ with identical bosons, only states
with even total angular momentum and positive parity ($J^{\pi}=0^{+},2^{+},
4^{+},\dots$) can contribute to the $^{7}$Li$(p,\alpha)^{4}$He and
$^{7}$Be$(n,\alpha)^{4}$He reactions; odd-$J$ and negative-parity states
($1^{+}$, $3^{+}$, $1^{-}$, $2^{-}$, \dots) are excluded by symmetry. In
Paper~I we found a twin $2^{+}$ doublet located just below the $p+{}^{7}$Li
threshold, and an
additional $2^{+}$ resonance at $E\approx 1.27$~MeV above the $^{7}$Be+$n$
threshold, while no $0^{+}$ resonance states appear near either the
$p+{}^{7}$Li or $^{7}$Be+$n$ thresholds. The different positions of these $2^{+}$ states relative to the two entrance
thresholds, together with the presence or absence of Coulomb repulsion in the
entrance channel, explain why the low--energy $S(E)$ behavior differs between
the two mirror reactions.

\begin{figure}[ptbh]
\begin{center}
\includegraphics[height=8.2857cm,width=13.7618cm]{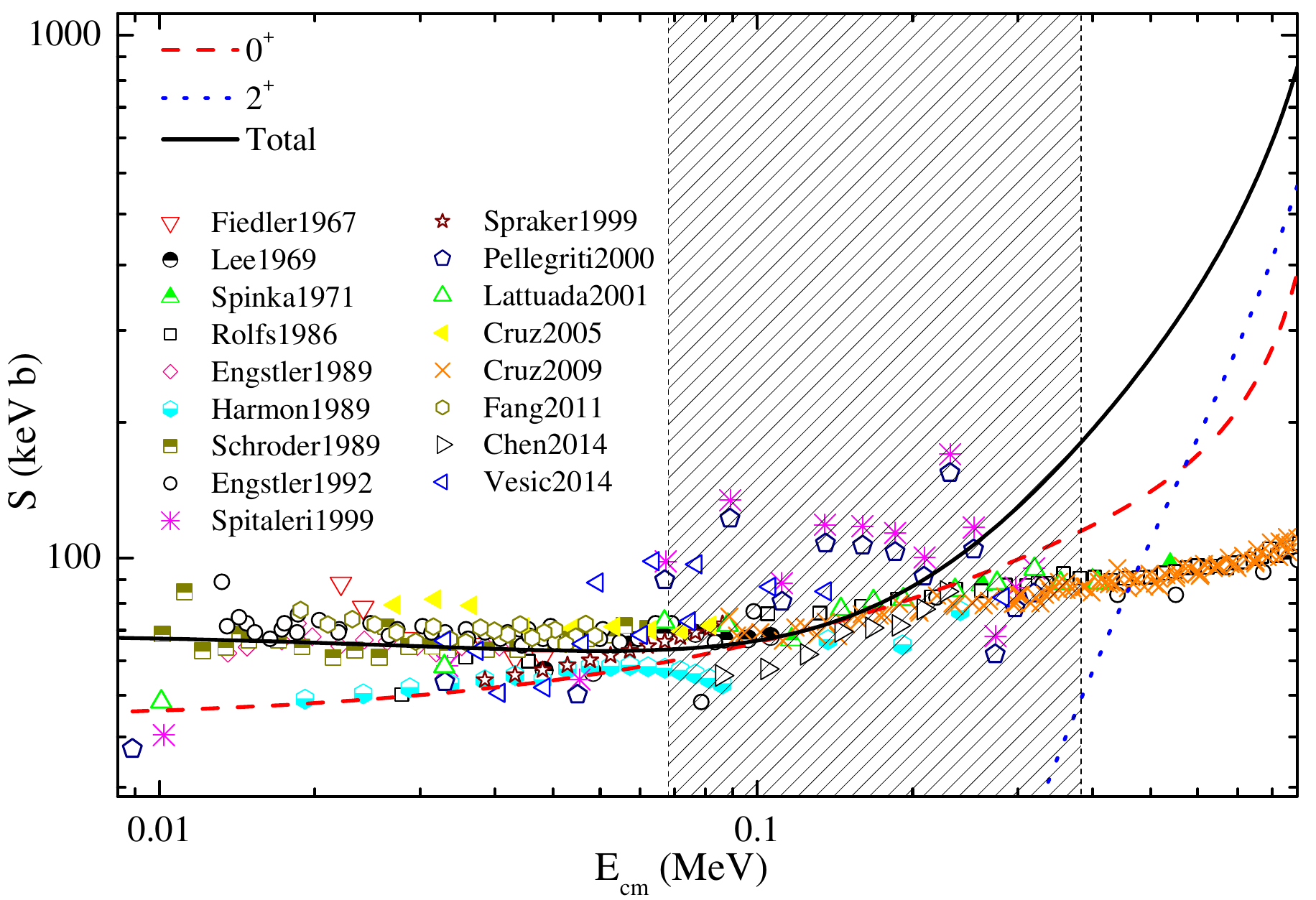}
\caption{Total and partial astrophysical $S$ factors for the reaction
$^{7}$Li$(p,\alpha)^{4}$He in the center-of-mass energy range
$E_{\mathrm{cm}} \lesssim 1$~MeV, calculated within the present model and
compared with the experimental data listed in
Table~\ref{Tab:SfactExpP7LiAA}. The dashed and dotted curves denote the
$J^{\pi}=0^{+}$ and $2^{+}$ contributions, respectively, and the solid curve
shows their sum. The shaded area indicates the Gamow window.}
\label{FIG:SfactorP7LiAAS}%
\end{center}
\end{figure}
Figure~\ref{FIG:SfactorP7LiAAS} shows the astrophysical $S$ factor for
$^{7}$Li$(p,\alpha)^{4}$He, calculated within the present model and compared
with the experimental data summarized in Table~\ref{Tab:SfactExpP7LiAA}. 
The dashed red, dotted blue, and solid black curves represent the $0^{+}$,
$2^{+}$, and total contributions, respectively; the $4^{+}$ contribution is
negligible over the energy range displayed and is not shown. The $0^{+}$
partial wave clearly dominates, while the $2^{+}$ component provides a
noticeable correction only at very low ($E_{\rm cm}\lesssim 100$~keV) and
relatively high ($E_{\rm cm}\gtrsim 500$~keV) energies. In our microscopic
spectrum there are no $0^{+}$ resonances close to the $p+{}^{7}$Li threshold,
and the only nearby structure is the upper member of the twin $2^{+}$ states,
located about 200~keV below that threshold. Consequently, the total $S(E)$
varies smoothly with energy and shows no narrow structures in the low--energy
region: the $0^{+}$ component sets the overall scale, while the subthreshold
$2^{+}$ resonance induces only a gentle curvature as $E\to 0$. The calculated
$S$ factor is in good agreement with the experimental data, particularly for
$E_{\rm cm}\lesssim 200$~keV.

\begin{figure}[ptbh]
\begin{centering}
\includegraphics[height=8.7887cm,width=14.5988cm]{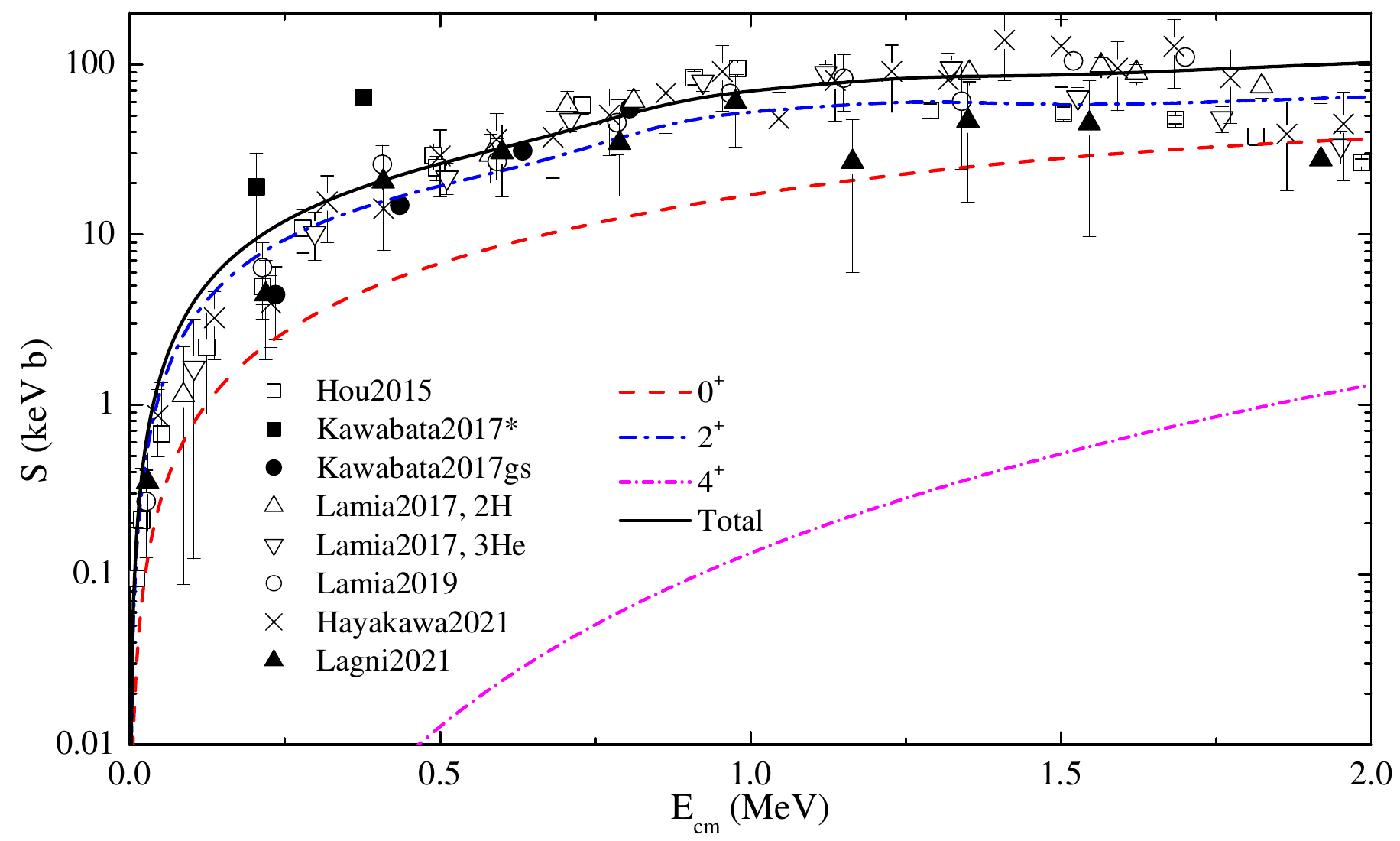}
\caption{Total and partial astrophysical $S$ factors for the reaction $^{7}$Be$(n,\alpha)^{4}$He, calculated within the present model and compared
with the experimental data summarized in Table~\ref{Tab:Exp7BeNAA}. All experimental data sets except those of Kawabata \textit{et al.} correspond to reactions on $^{7}$Be in its ground state; the points labeled “Kawabata2017gs” and “Kawabata2017*” represent reactions on the ground and first excited states, respectively. The dashed curves show the $J^{\pi}=0^{+}$, $2^{+}$, and $4^{+}$ partial contributions, and the solid curve is their sum.}
\label{FIG:SfactoN7BeAAS}%
\end{centering}
\end{figure}

The total and partial astrophysical $S$ factors for $^{7}$Be$(n,\alpha)^{4}$He
are shown in Fig.~\ref{FIG:SfactoN7BeAAS}, together with the experimental data
summarized in Table~\ref{Tab:Exp7BeNAA}. Compared with the mirror
$^{7}$Li$(p,\alpha)^{4}$He channel, the $^{7}$Be$(n,\alpha)^{4}$He reaction
exhibits a different hierarchy of partial contributions: the $2^{+}$ wave
dominates over the whole energy range, followed by the $0^{+}$ component. The
$4^{+}$ contribution is small over most of the range but becomes more visible
than in the proton--induced reaction at higher energies, while remaining
subdominant to the $2^{+}$ and $0^{+}$ parts. The absence of Coulomb repulsion in the entrance channel, together with the
proximity of a higher $2^{+}$ resonance located at $E\approx 1.27$~MeV above the $^{7}$Be+$n$ threshold (and thus closer to
this threshold than to the $p+{}^{7}$Li threshold), enhances the weight of the
$2^{+}$ partial wave and leads to a low--energy $S(E)$ that differs from the
almost flat behavior of the mirror $p+{}^{7}$Li reaction.
 Overall, the calculated total $S$ factor reproduces the
magnitude and energy dependence of the experimental data for
$^{7}$Be$(n,\alpha)^{4}$He within the quoted uncertainties.

\begin{figure}[ptbh]
\begin{center}
\includegraphics[height=8.7887cm,width=14.5988cm]{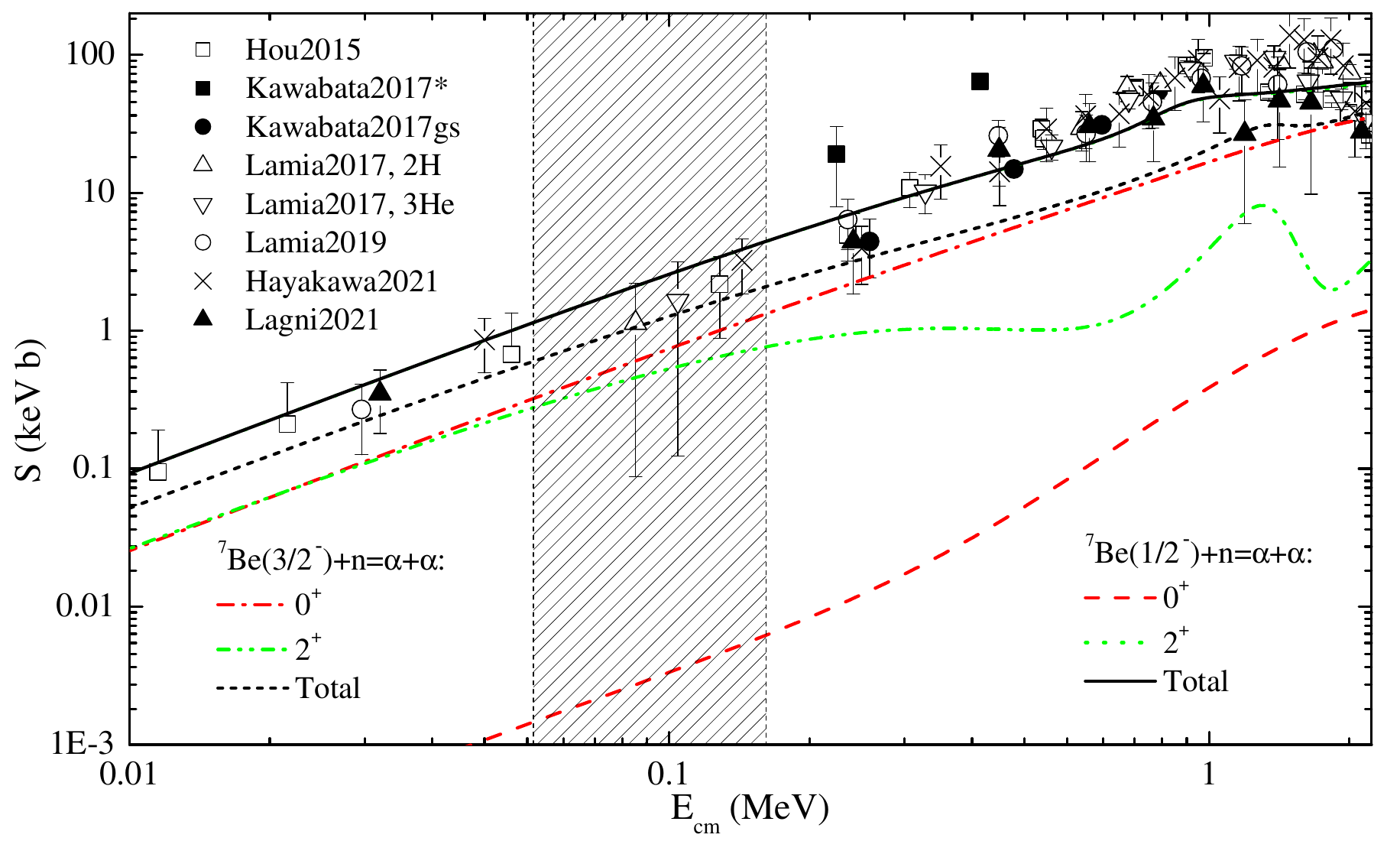}
\caption{Low-energy astrophysical $S$ factors for the reaction
$^{7}$Be$(n,\alpha)^{4}$He, calculated within the present model separately for
initial $^{7}$Be in the ground ($3/2^{-}$) and first excited ($1/2^{-}$)
states, and compared with the experimental data summarized in
Table~\ref{Tab:Exp7BeNAA} (all non-Kawabata points correspond to
reactions on $^{7}$Be in its ground state, while ``Kawabata2017gs'' and
``Kawabata2017*'' denote the ground and first excited states, respectively). For each initial state, the red and green curves
denote the $0^{+}$ and $2^{+}$ contributions, respectively, and the black curve
gives their sum.}
\label{FIG:SfactoN7BeAAS0}%
\end{center}
\end{figure}

Figure~\ref{FIG:SfactoN7BeAAS0} further decomposes the calculated $S$ factors
into contributions from reactions on $^{7}$Be in its ground ($3/2^{-}$) and
first excited ($1/2^{-}$) states. For the $^{7}$Be($1/2^{-}$) initial state
the $2^{+}$ partial wave almost saturates the total $S$ factor, so that the
total curve nearly coincides with the $2^{+}$ contribution. In contrast, for
$^{7}$Be($3/2^{-}$)$(n,\alpha)^{4}$He the $0^{+}$ and $2^{+}$ components are
of comparable magnitude, and a pronounced enhancement due to the $2^{+}$
resonance is clearly visible in the $2^{+}$ curve. In our microscopic
calculation this $2^{+}$ state appears at $E\approx 1.27$~MeV above the
$^{7}$Be+$n$ threshold, in good agreement with the value adopted by
Tilley \textit{et al.}~\cite{2004NuPhA.745..155T} and somewhat higher than the
resonance energy $E\approx 0.99$~MeV reported by Hayakawa
\textit{et al.}~\cite{2021ApJ...915L..13H}. Both analyses require enhanced
$2^{+}$ contribution in this region, and the energy dependence of our $2^{+}$
component is qualitatively consistent with the behavior inferred from the data
of Ref.~\cite{2021ApJ...915L..13H}.

As a consequence of this partial--wave pattern, our calculation predicts that
the $^{7}$Be($1/2^{-}$) $(n,\alpha)^{4}$He channel gives a larger contribution
than the $^{7}$Be($3/2^{-}$)$(n,\alpha)^{4}$He channel over most of the energy
range. This trend is consistent with the state--resolved cross sections
extracted by Kawabata \textit{et al.}~\cite{2017PhRvL.118e2701K}, whose data
also indicate a larger cross section for reactions on the first excited state
(the points labeled ``Kawabata2017*'' and ``Kawabata2017gs'' in
Fig.~\ref{FIG:SfactoN7BeAAS0}). The sum of the two initial--state
contributions reproduces the total $S$ factor shown in
Fig.~\ref{FIG:SfactoN7BeAAS}, in good agreement with the experimental data.
In this sense, the relative weight of the ground and first excited states of
$^{7}$Be emerges as a sensitive observable: the present microscopic model
provides a definite prediction for this ratio, which can be further tested by
future experiments with improved control over the $^{7}$Be beam composition
and state selectivity.

\subsection{Charge--exchange reaction $^{7}$Be$(n,p)^{7}$Li}

\begin{figure}[ptbh]
\begin{center}
\includegraphics[height=8.2527cm,width=13.7091cm]{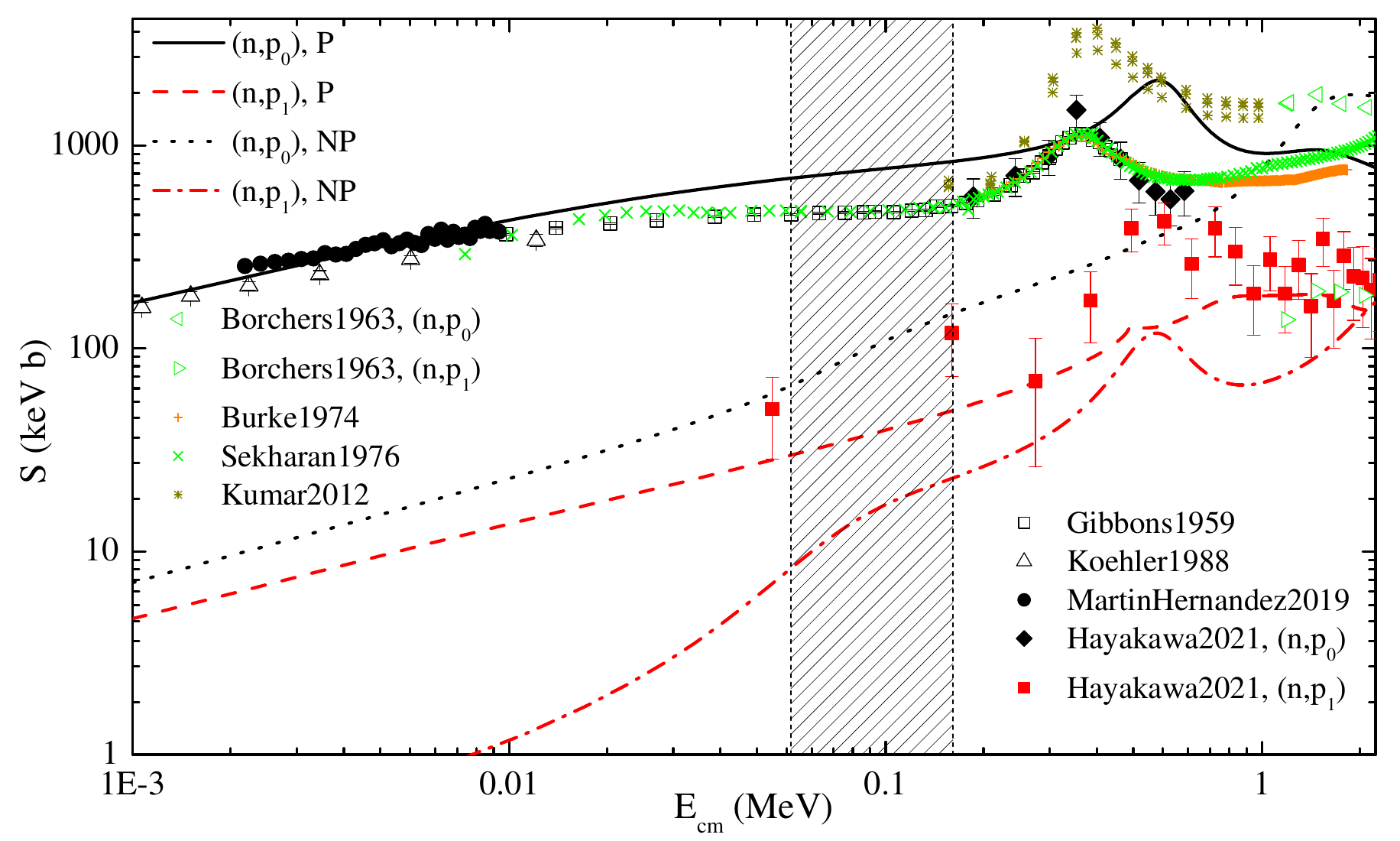}
\caption{Astrophysical $S$ factors for the reaction $^{7}$Be$(n,p)^{7}$Li,
calculated within the present model and compared with experimental data
summarized in Table~\ref{Tab:Exp7BeNP7Li}. The $(n,p_{0})$ and $(n,p_{1})$
channels correspond to $^{7}$Li in the ground ($3/2^{-}$) and first excited
($1/2^{-}$) states, respectively. The labels ``P'' and ``NP'' denote
calculations with and without cluster polarization. All experimental data
sets, except those of Borchers and Hayakawa, correspond to the $(n,p_{0})$
channel.}
\label{FIG:SfactoN7BeP7Li}%
\end{center}
\end{figure}

The astrophysical $S$ factors for $^{7}$Be$(n,p)^{7}$Li, calculated within
the present model with and without cluster polarization, are shown in
Fig.~\ref{FIG:SfactoN7BeP7Li} and compared with the experimental data
summarized in Table~\ref{Tab:Exp7BeNP7Li}. 
The figure displays both
$^{7}$Be$(n,p_{0})^{7}$Li and $^{7}$Be$(n,p_{1})^{7}$Li channels, each shown
for calculations with (P) and without (NP) cluster polarization. Our results
exhibit a pronounced enhancement in the $^{7}$Be$(n,p_{0})^{7}$Li $S$ factor
near $E_{\mathrm{cm}}\simeq 0.5$~MeV. This structure is generated by the
higher member of the twin $3^{+}$ resonance doublet, which in our model lies
about 0.53~MeV above the $^{7}$Be+$n$ threshold. In the
$^{7}$Be$(n,p_{1})^{7}$Li channel the peak just below 0.5~MeV is produced by
a $1^{-}$ resonance located at $E\approx 0.45$~MeV. The experimental data
show a similar pattern: both channels develop maxima in this energy region,
but the peak in the measured $S$ factor for $^{7}$Be$(n,p_{0})^{7}$Li is
slightly shifted toward lower energies, consistent with the fact that our
calculation places the $3^{+}$ resonance about 200~keV higher than the
experimental value.

As demonstrated in Paper~I, cluster polarization plays a critical role in the
formation of the $1^{+}$, $2^{+}$, $3^{+}$, and $4^{+}$ twin resonance
states. Figure~\ref{FIG:SfactoN7BeP7Li} confirms that it also has a
substantial impact on the astrophysical $S$ factor for
$^{7}$Be$(n,p)^{7}$Li, especially at low energies. Calculations without
cluster polarization (NP) underestimate the $S$ factor already at very low
energies and throughout the Gamow window and place the $(n,p_{0})$ peak at
too high an energy. Including cluster polarization (P) enhances the dynamical
coupling between entrance and exit channels, increases the $S$ factor at very
low energies, and shifts the $(n,p_{0})$ maximum toward the experimental
value, leading to good overall agreement with the data.
\begin{figure}[ptbh]
\begin{center}
\includegraphics[width=0.49\textwidth]{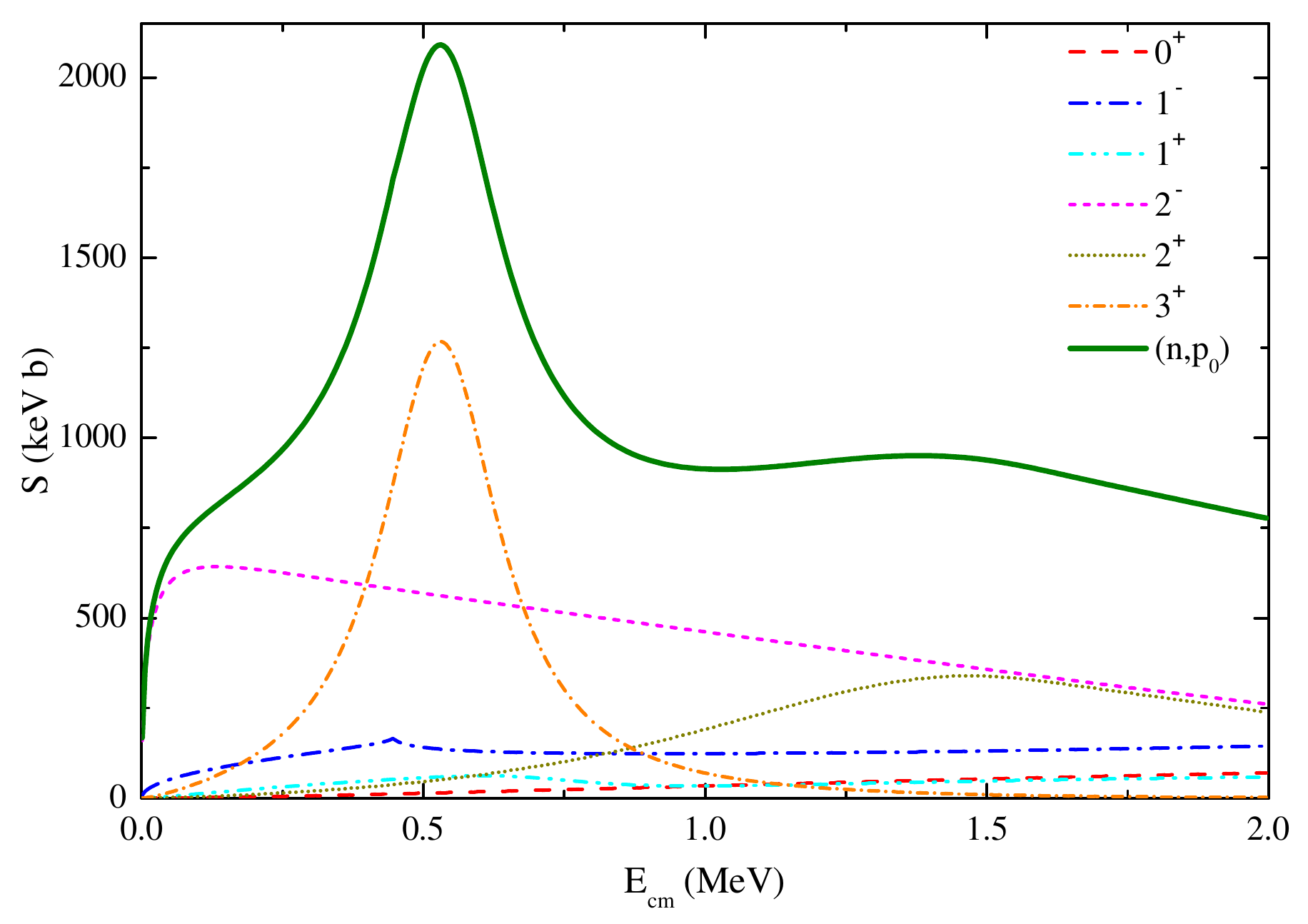}
\includegraphics[width=0.49\textwidth]{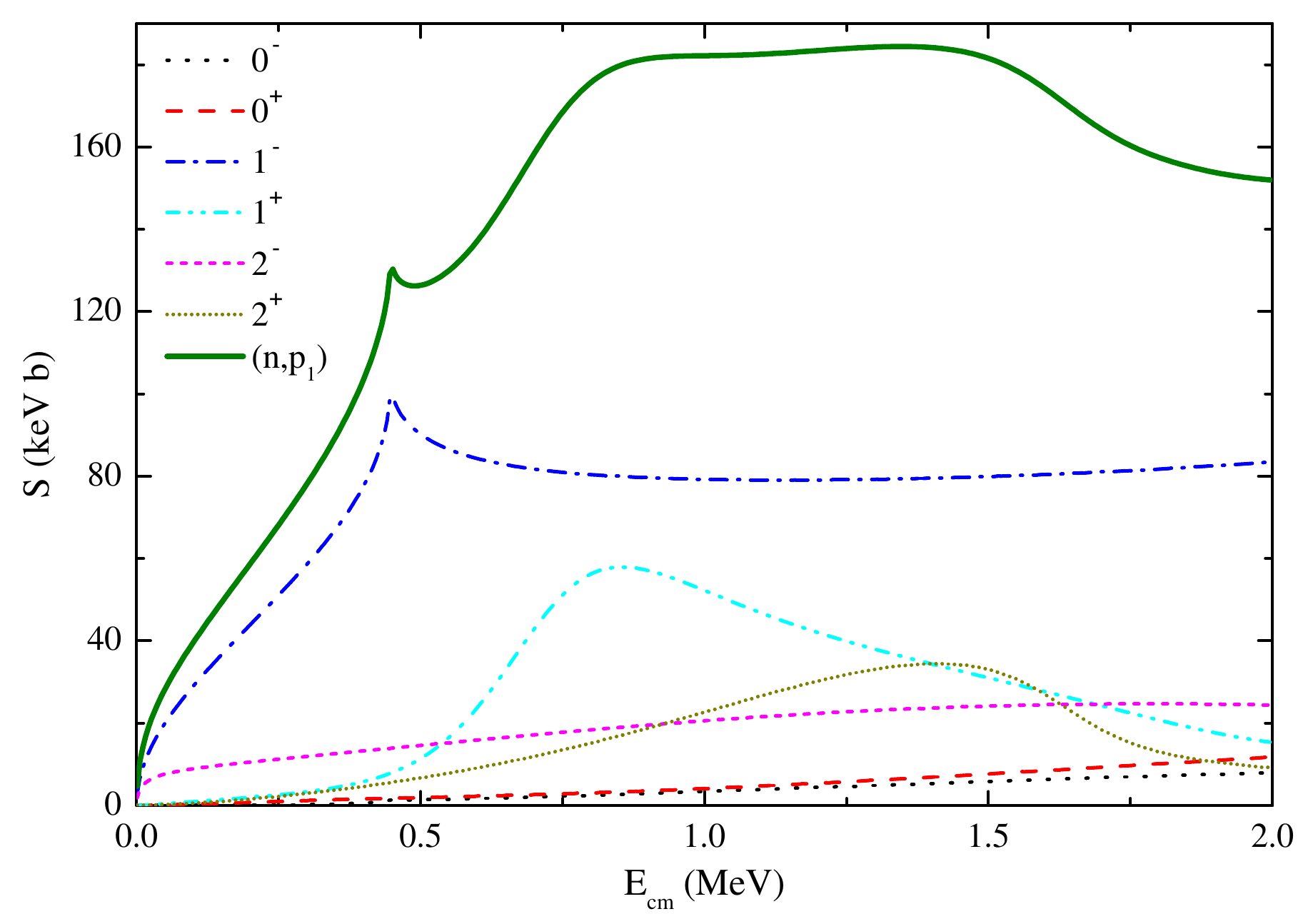}
\caption{Contribution of states with the different values of the total angular
momentum $J$  to the total astrophysical S-factor of the reaction $^{7}%
$Be($n$,$p$)$^{7}$Li. The $(n,p_{0})$ and $(n,p_{1})$ channels correspond to $^{7}$Li in the ground ($3/2^-$) and first excited ($1/2^-$) states, respectively. }
\label{Fig:SFactorsNPvsJ}%
\end{center}
\end{figure}
Further insight into the charge--exchange mechanism is provided by the
decomposition into total angular momentum $J$, shown in
Fig.~\ref{Fig:SFactorsNPvsJ}. As discussed above, the $\alpha+\alpha$ exit channel is restricted to even-$J$,
positive-parity states. By contrast, $^{7}$Be$(n,p)^{7}$Li can proceed
through both parities and a wider set of $J$ values.
 In the $^{7}$Be$(n,p_{0})^{7}$Li channel, the $3^{+}$ resonance dominates the peak
region just above 0.5~MeV, whereas a $2^{-}$ state located very close to the
$^{7}$Be+$n$ threshold governs the behavior at both lower and higher
energies. In contrast, the $^{7}$Be$(n,p_{1})^{7}$Li channel is dominated by
the $1^{-}$ state over the entire energy range: the corresponding $1^{-}$
resonance at $E\approx 0.45$~MeV fixes the position of the peak in the
$(n,p_{1})$ $S$ factor and provides the main contribution away from the
maximum.

A comparison of the astrophysical $S$ factors for the two neutron--induced
reactions on $^{7}$Be, $^{7}$Be$(n,\alpha)^{4}$He and
$^{7}$Be$(n,p)^{7}$Li, shows that in the low--energy region the charge--exchange
channel $^{7}$Be$(n,p)^{7}$Li clearly dominates over the $\alpha$--emission
channel. Thus, at astrophysically relevant energies neutron interactions with
$^{7}$Be predominantly lead to the production of $^{7}$Li.

\subsection{Deuteron--induced reactions on $^{6}$Li}

We now turn to deuteron--induced reactions on $^{6}$Li, where all three
channels share a common sensitivity to the $^{6}$Li+$d$ threshold and to
broad $2^{+}$ resonance in $^{8}$Be just below this threshold. As discussed in
Paper~I, the present model places the $^{6}$Li+$d$ threshold somewhat too
high compared to experiment, which leads to a systematic underestimation of
the calculated $S$ factors for the $^{6}$Li$(d,\alpha)^{4}$He,
$^{6}$Li$(d,p)^{7}$Li, and $^{6}$Li$(d,n)^{7}$Be reactions at low
entrance--channel energies. Since all three exit channels originate from the
same $^{6}$Li+$d$ entrance configuration, we therefore examine not only the
total $S(E)$ but also the $J^{\pi}$ decomposition, in order to see which
partial waves dominate the calculated $S$ factors in each exit channel.

\begin{figure}[ptbh]
\begin{center}
\includegraphics[height=8.5558cm,width=14.043cm]{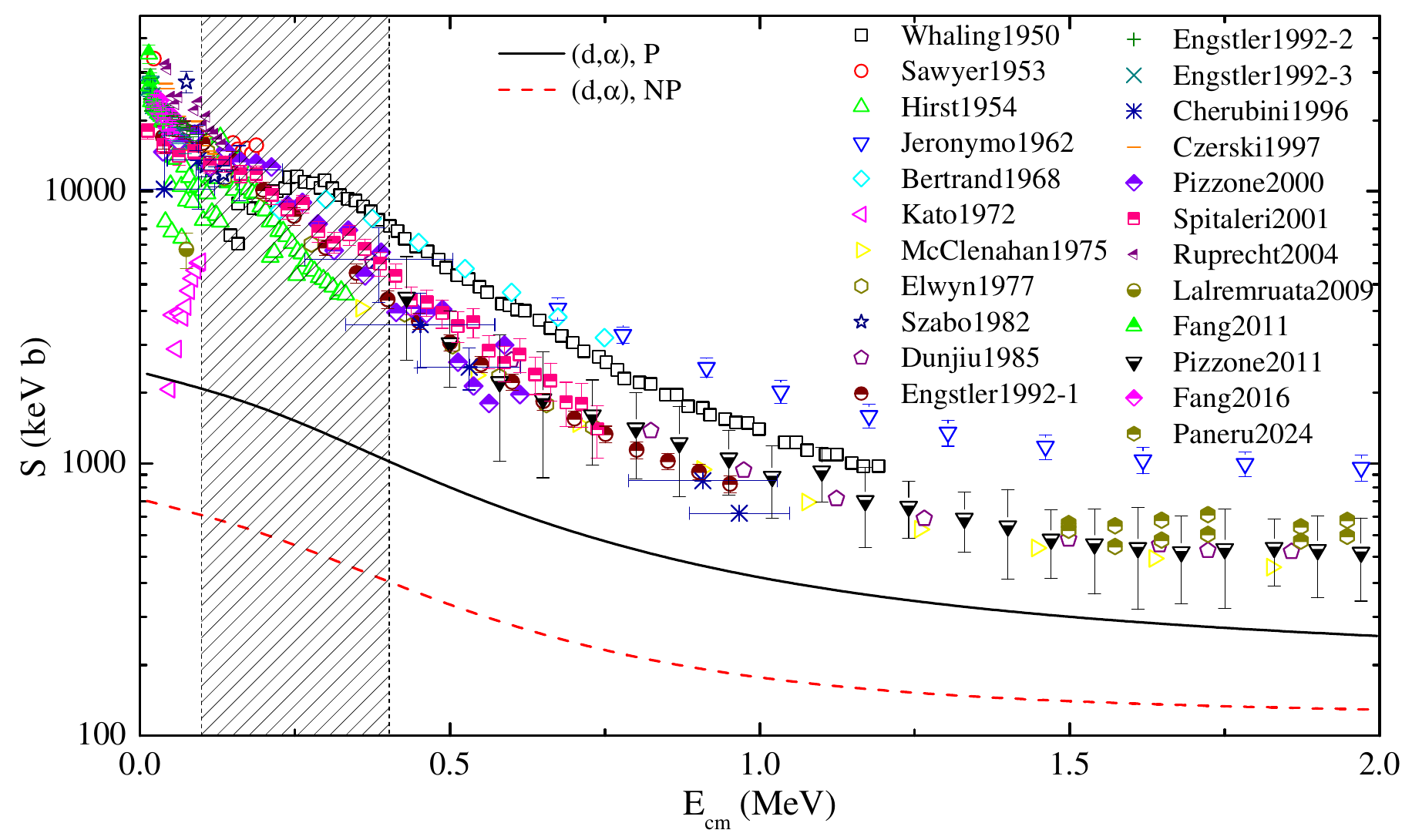}%
\caption{Astrophysical $S$ factor for the reaction $^{6}$Li$(d,\alpha)^4$He,
calculated within the present model (lines) and compared with experimental
data summarized in Table \ref{Tab:SfactExp62AA}. P and NP denote calculations
with and without cluster polarization.}
\label{FIG:SfactorD6LiAA}%
\end{center}
\end{figure}

Figure~\ref{FIG:SfactorD6LiAA} shows the astrophysical $S$ factor for
$^{6}$Li$(d,\alpha)^{4}$He, calculated within the present model and compared
with the experimental data summarized in Table~\ref{Tab:SfactExp62AA} of
Appendix~\ref{Sec:Exper}. Owing to the shifted position of the
$^{6}$Li+$d$ threshold, the calculated $S$ factor remains lower than most
experimental values in the range $E_{\rm cm}\lesssim 1$~MeV. Including
cluster polarization increases the $S$ factor and improves agreement with the
data, although the enhancement is not sufficient to fully reproduce the
experimental magnitudes. At higher energies the theoretical results
gradually converge toward the experimental trend.

\begin{figure}[ptbh]
\begin{center}
\includegraphics[height=8.5558cm,width=14.043cm]{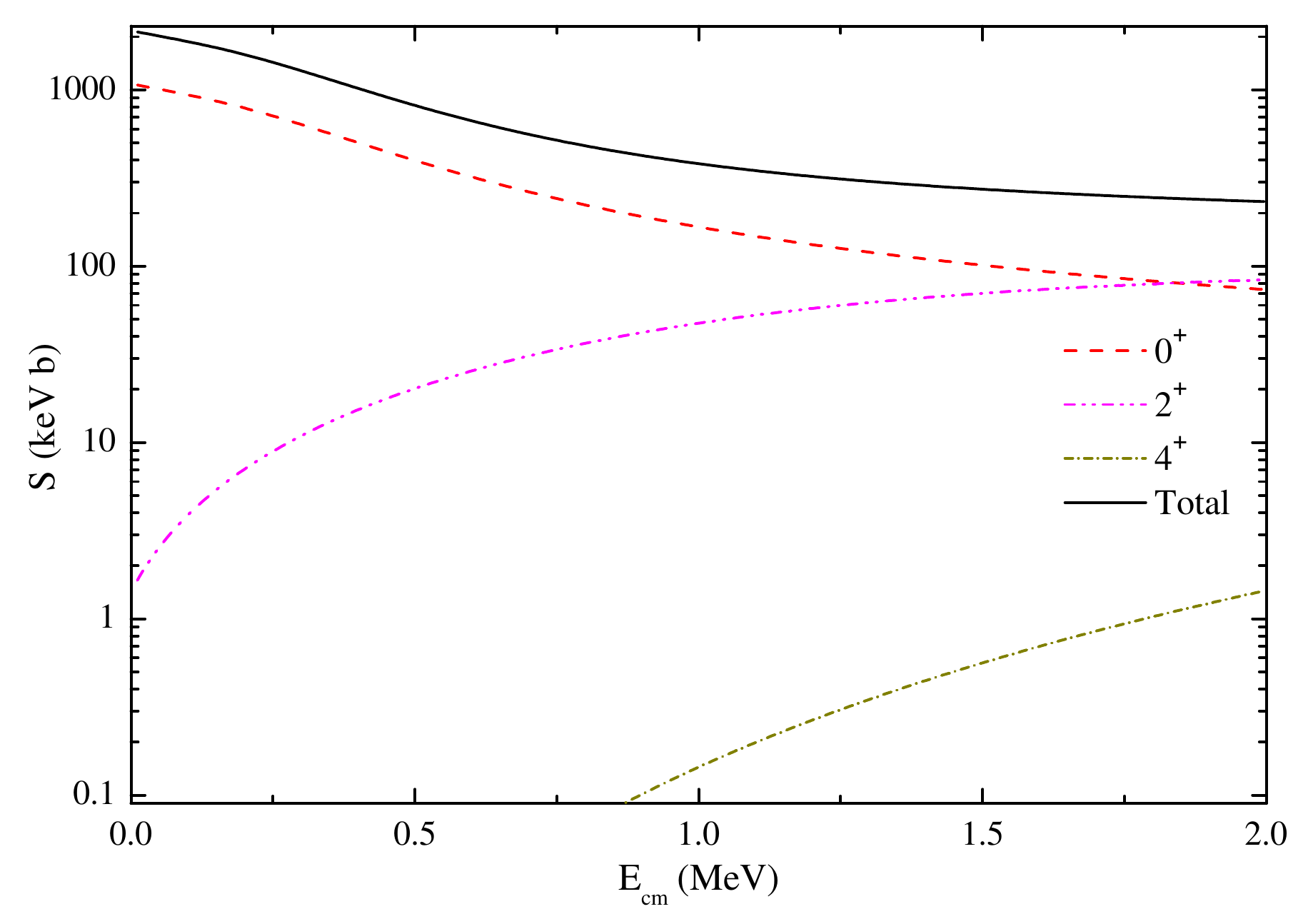}%
\caption{Contributions of states with different total angular momenta $J$ to
the astrophysical $S$ factor for the reaction $^{6}$Li$(d,\alpha)^4$He,
calculated within the present model.}
\label{FIG:SfactorD6LiLS}%
\end{center}
\end{figure}

The partial $S$ factors shown in Fig.~\ref{FIG:SfactorD6LiLS} indicate that
the $0^{+}$ state provides the dominant contribution to the total
$^{6}$Li$(d,\alpha)^4$He $S$ factor below $E_{\rm cm}\approx 1.5$~MeV; above
this energy the $2^{+}$ contribution becomes comparable, while the $4^{+}$
component remains small over the entire range displayed. Thus, as in the
$\alpha$--emission reactions on $A=7$ nuclei, the $\alpha+\alpha$ exit
channel is governed primarily by $0^{+}$ and $2^{+}$ waves, with the missing
broad $2^{+}$ resonance near the $^{6}$Li+$d$ threshold in the present
implementation largely responsible for the underestimated $S$ factor at low
energies.

\begin{figure}[ptbh]
\begin{center}
\includegraphics[height=8.4526cm,width=14.043cm]{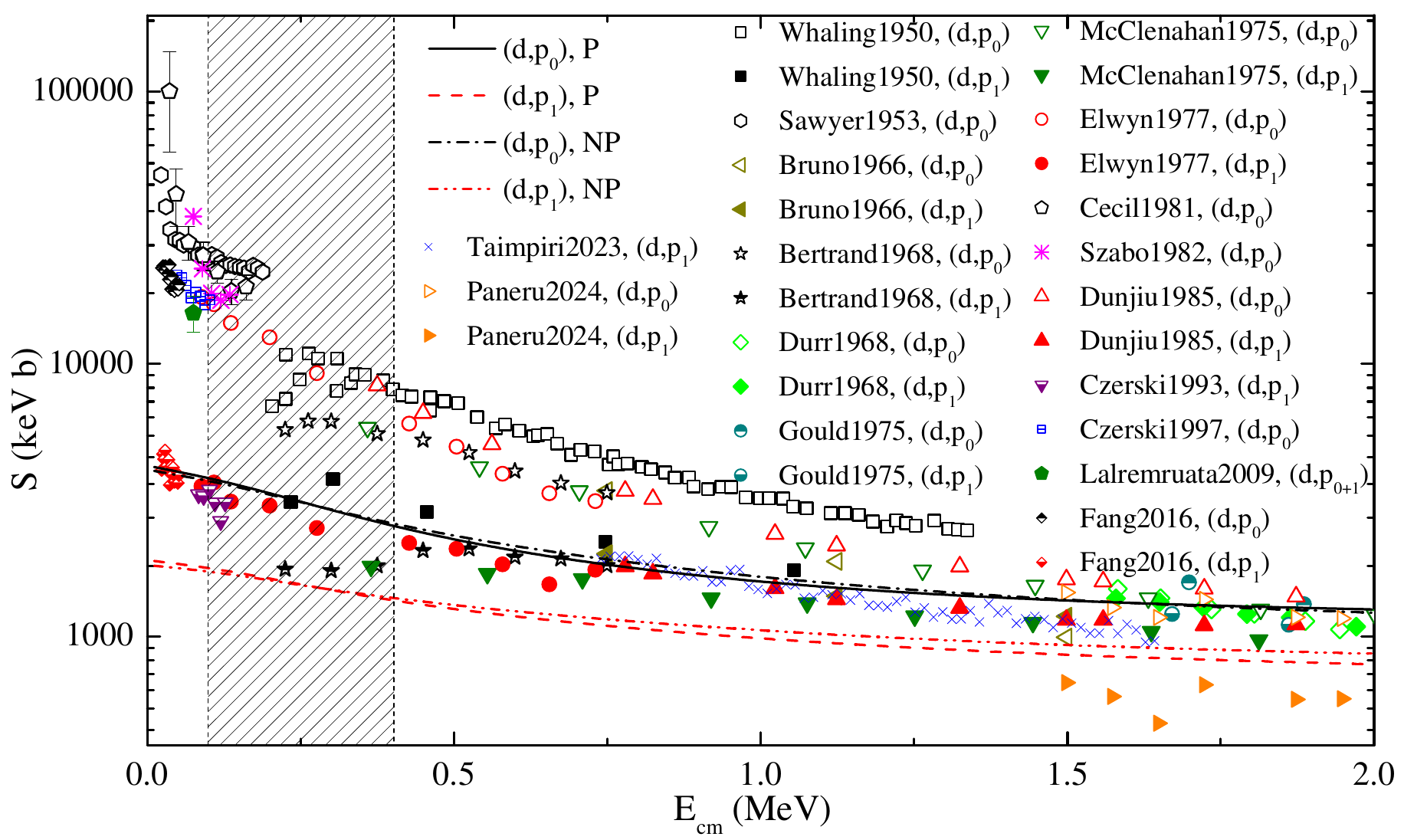}%
\caption{Astrophysical $S$ factors for the reaction $^{6}$Li$(d,p)^{7}$Li,
calculated within the present model (lines) and compared with experimental
data summarized in Table~\ref{Tab:SfactExp62P7Li}. The $(d,p_{0})$ and
$(d,p_{1})$ channels correspond to population of $^{7}$Li in the ground
($3/2^{-}$) and first excited ($1/2^{-}$) states, respectively. The labels
``P'' and ``NP'' denote calculations with and without cluster polarization,
respectively.}
\label{FIG:SfactorD6LiP7Li}%
\end{center}
\end{figure}

\begin{figure}[ptbh]
\begin{center}
\includegraphics[height=8.4526cm,width=14.043cm]{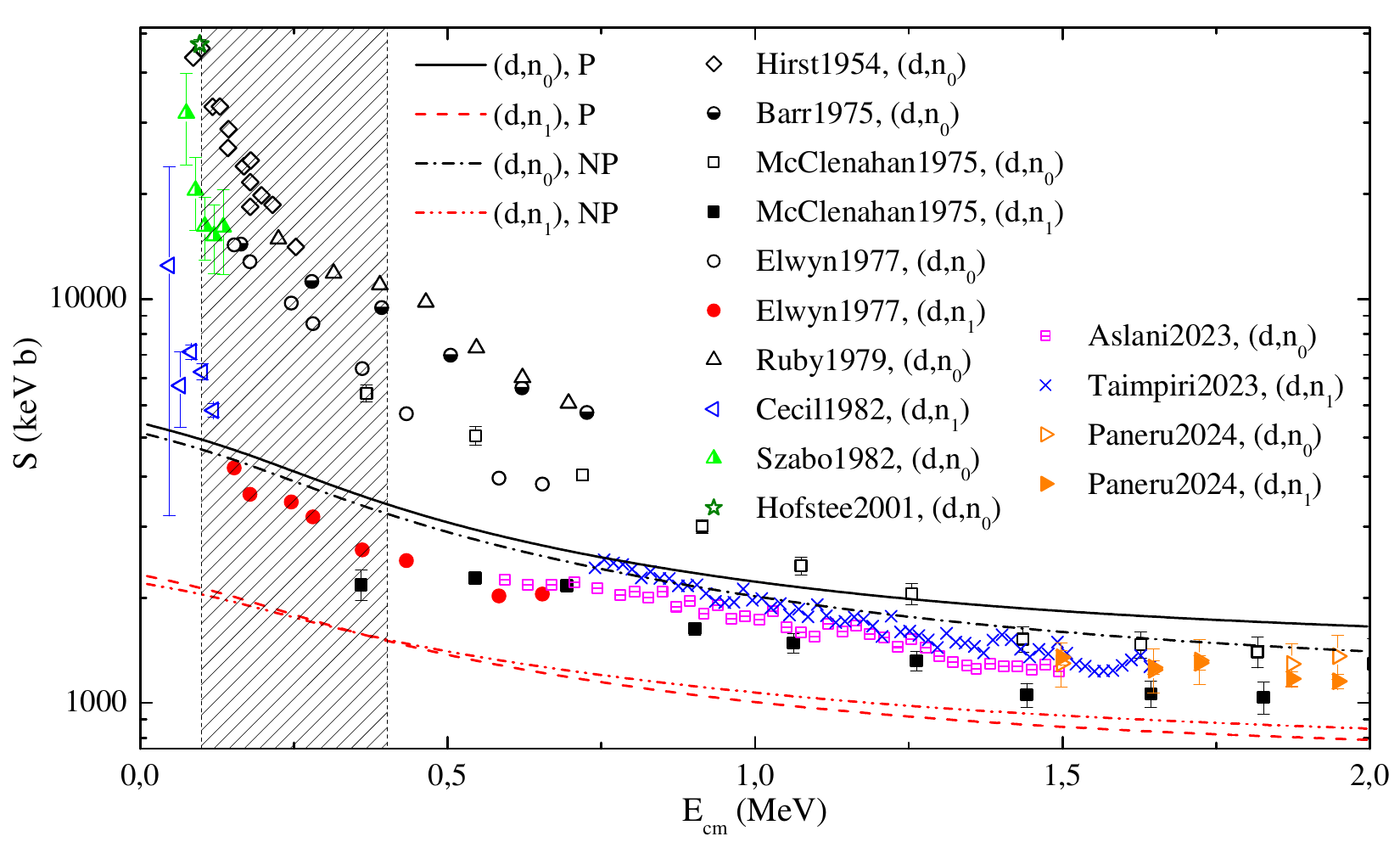}%
\caption{Astrophysical $S$ factors for the reaction $^{6}$Li$(d,n)^{7}$Be,
calculated within the present model (lines) and compared with experimental
data summarized in Table~\ref{Tab:SfactExp62N7Be}. The $(d,n_{0})$ and
$(d,n_{1})$ channels correspond to $^{7}$Be in the ground ($3/2^{-}$) and
first excited ($1/2^{-}$) states, respectively. The labels ``P'' and ``NP''
denote calculations with and without cluster polarization, respectively.}
\label{FIG:SfactorD6LiN7Be}%
\end{center}
\end{figure}

The astrophysical $S$ factors for the $^{6}$Li$(d,p)^{7}$Li and
$^{6}$Li$(d,n)^{7}$Be reactions are shown in Figs.~\ref{FIG:SfactorD6LiP7Li}
and \ref{FIG:SfactorD6LiN7Be}, respectively, with the corresponding
experimental data summarized in Tables~\ref{Tab:SfactExp62P7Li} and
\ref{Tab:SfactExp62N7Be}. In Fig.~\ref{FIG:SfactorD6LiP7Li} most experimental
data for $^{6}$Li$(d,p_{1})^{7}$Li lie very close to our calculated
$^{6}$Li$(d,p_{0})^{7}$Li curve, whereas the experimental $S$ factors for
$^{6}$Li$(d,p_{0})^{7}$Li itself exceed the theoretical predictions for
$E_{\rm cm}\lesssim 1$~MeV. A similar pattern is seen for
$^{6}$Li$(d,n)^{7}$Be in Fig.~\ref{FIG:SfactorD6LiN7Be}, where the
calculated $S$ factors reproduce the shapes of the data but underestimate
their absolute scale at low energies.

As seen in Figs.~\ref{FIG:SfactorD6LiP7Li} and \ref{FIG:SfactorD6LiN7Be},
cluster polarization has little effect on the $^{6}$Li$(d,p)^{7}$Li and
$^{6}$Li$(d,n)^{7}$Be $S$ factors, in contrast to the
$^{6}$Li$(d,\alpha)^{4}$He and, especially, the $^{7}$Be$(n,p)^{7}$Li
channels, where its impact is much more pronounced. In these deuteron--induced
$(d,p)$ and $(d,n)$ channels the remaining discrepancy with experiment at
$E_{\rm cm}\lesssim 1$~MeV is therefore mainly attributable to the misplaced
$^{6}$Li+$d$ threshold rather than to the treatment of cluster polarization.

\begin{figure}[ptbh]
\begin{center}
\includegraphics[width=0.49\textwidth]{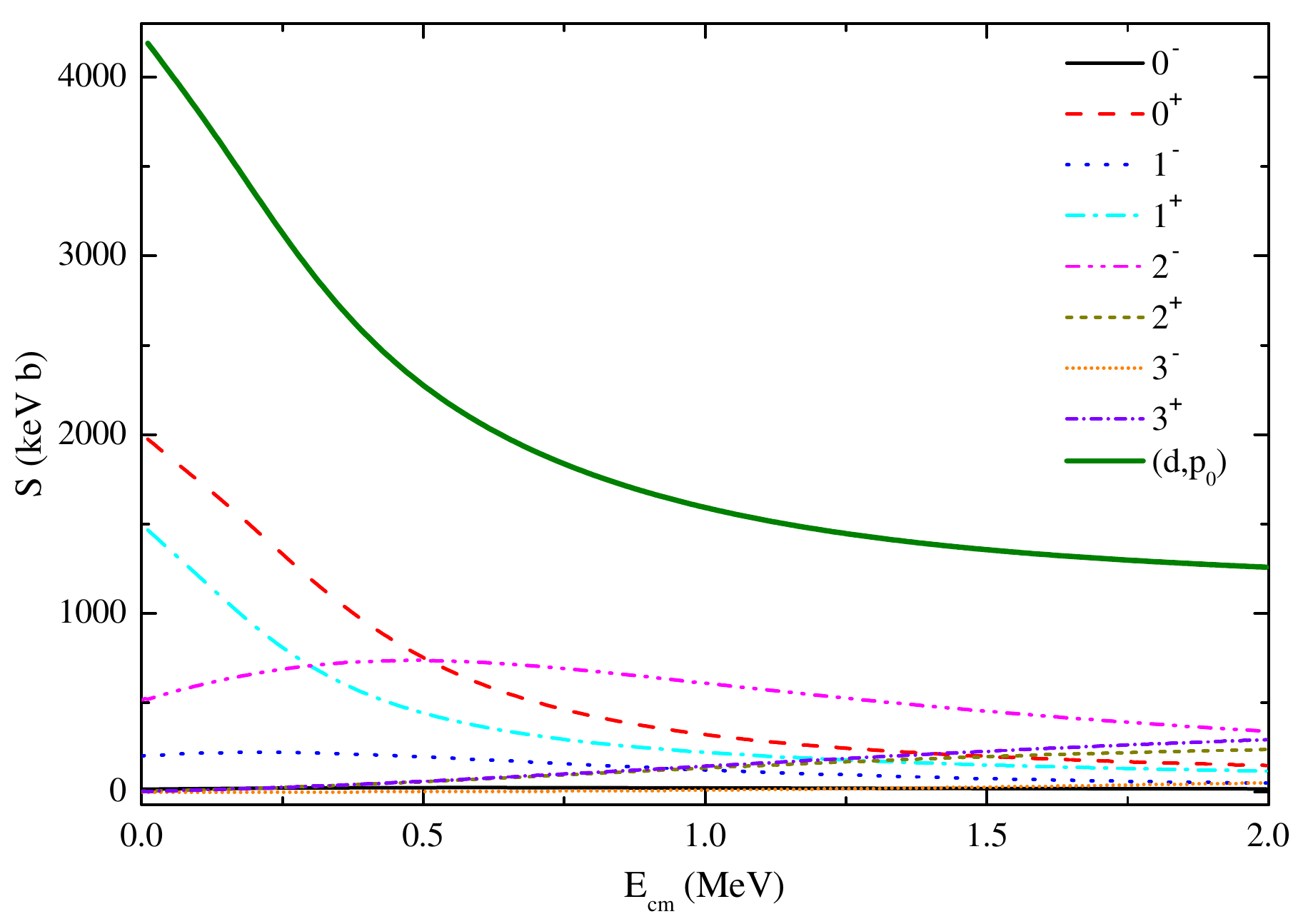}
\includegraphics[width=0.49\textwidth]{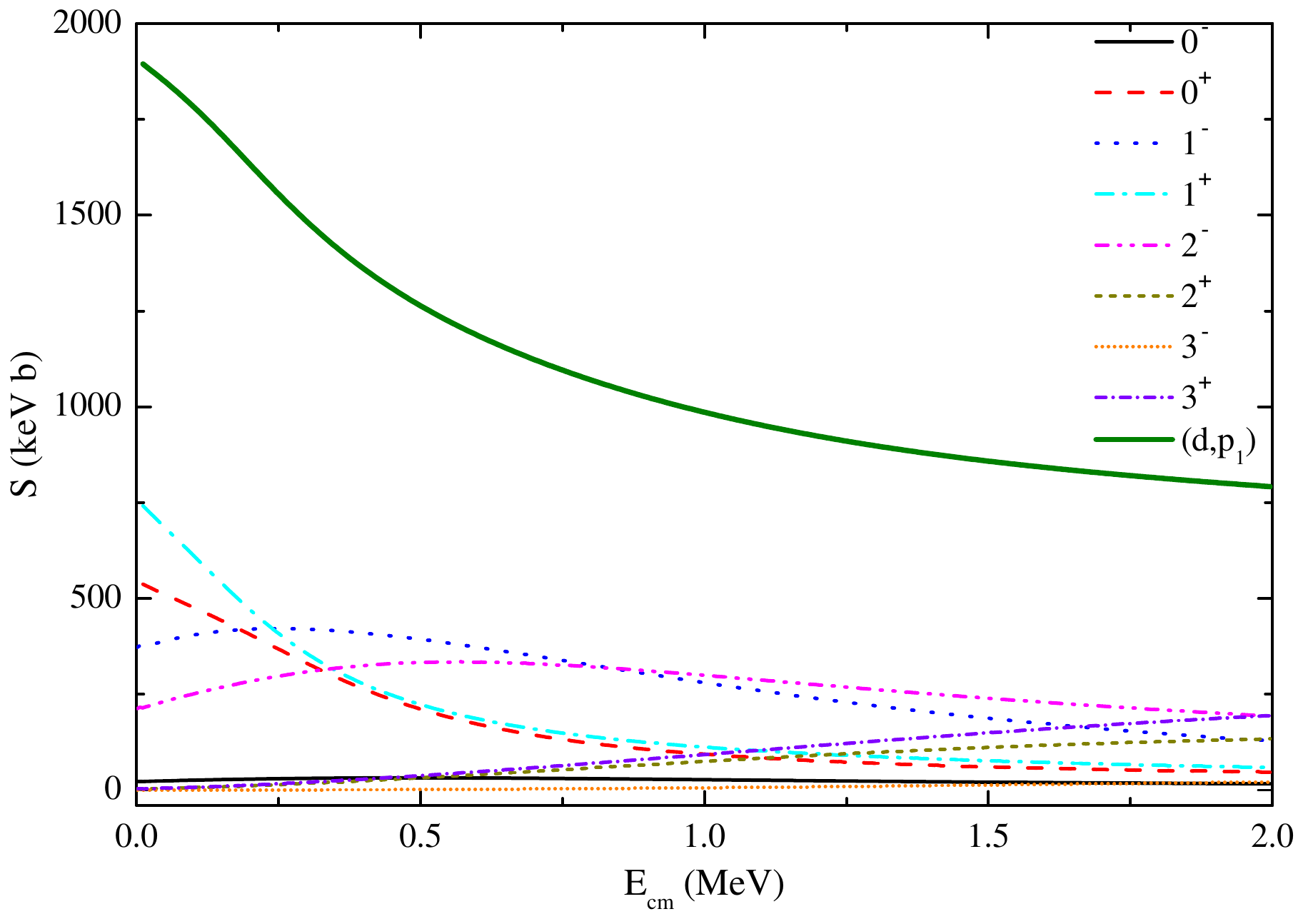}
\caption{Contributions of states with different total angular momenta $J$ to
the astrophysical $S$ factor of the $^{6}$Li$(d,p)^{7}$Li reaction. The
$(d,p_{0})$ and $(d,p_{1})$ channels correspond to $^{7}$Li in the ground
($3/2^-$) and first excited ($1/2^-$) states, respectively.}
\label{Fig:SFactors6LiDP7LiLS}%
\end{center}
\end{figure}

\begin{figure}[ptbh]
\begin{center}
\includegraphics[width=0.49\textwidth]{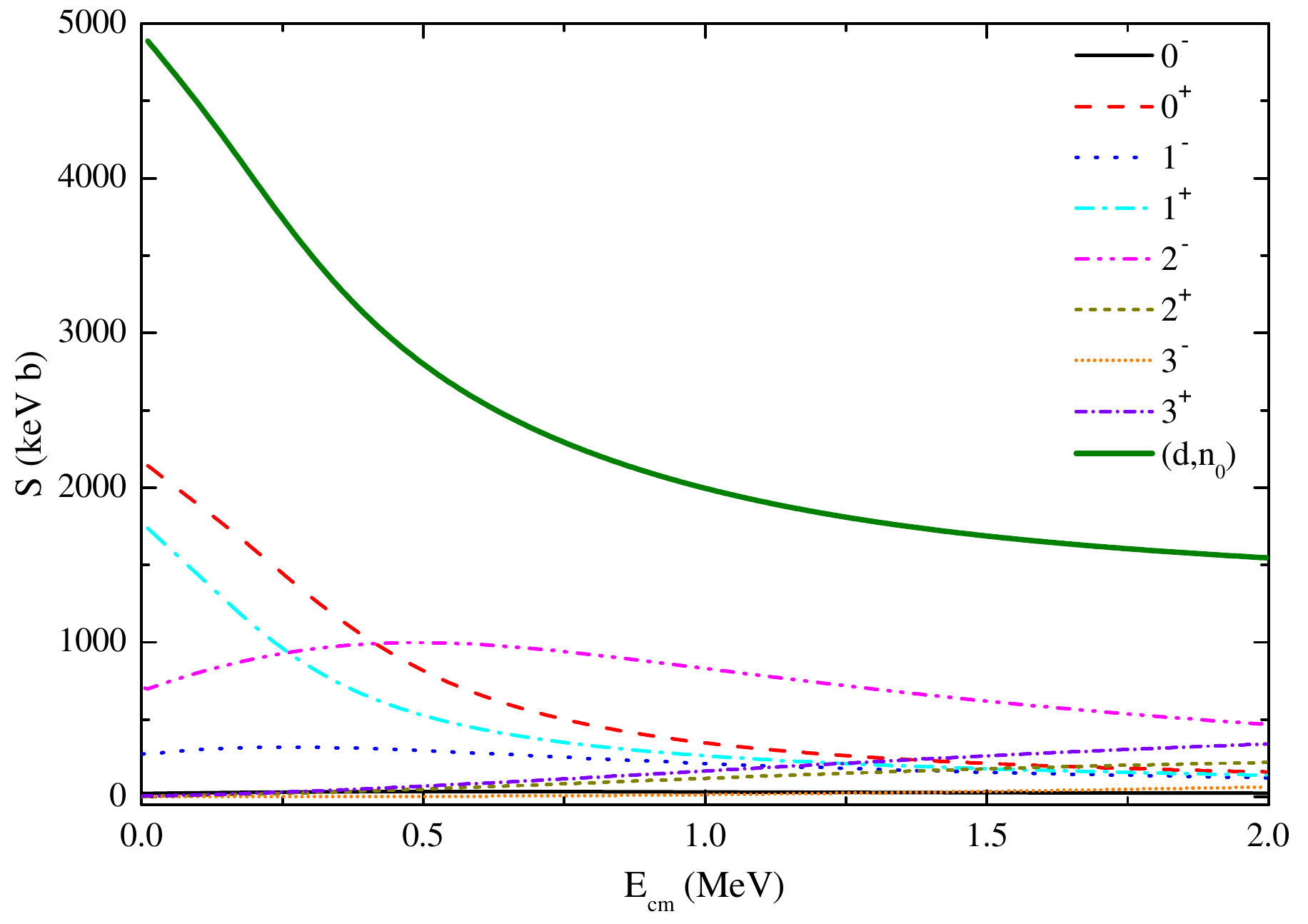}%
\includegraphics[width=0.49\textwidth]{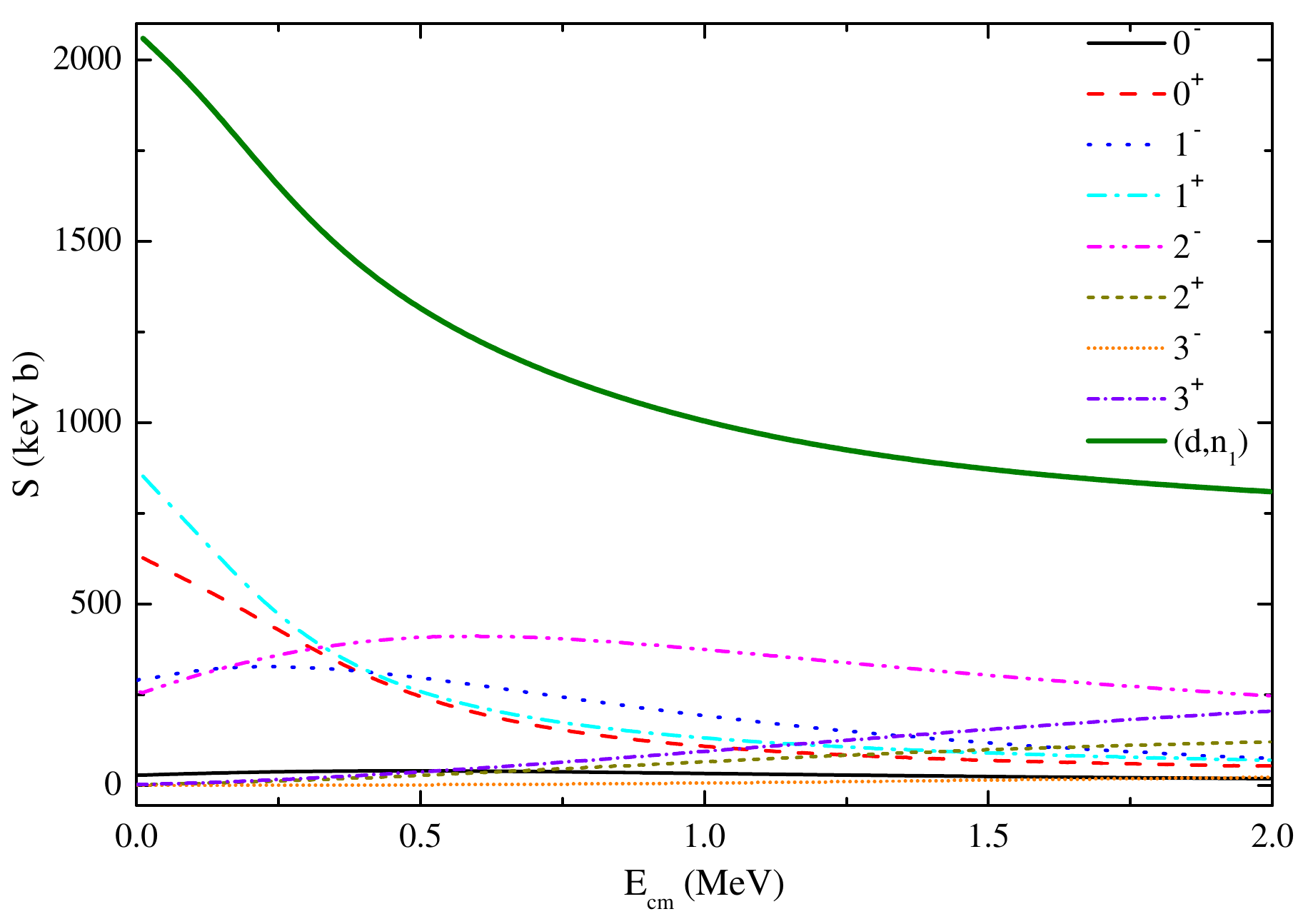}
\caption{Contributions of states with different total angular momenta $J$ to
the astrophysical $S$ factor of the $^{6}$Li$(d,n)^{7}$Be reaction. The
$(d,n_{0})$ and $(d,n_{1})$ channels correspond to $^{7}$Be in the ground
($3/2^-$) and first excited ($1/2^-$) states, respectively.}
\label{Fig:SFactors6LiDN7BeLS}%
\end{center}
\end{figure}

The $J^{\pi}$ decomposition provides additional insight into the structure of
these charge--exchange channels. Figures~\ref{Fig:SFactors6LiDP7LiLS} and
\ref{Fig:SFactors6LiDN7BeLS} show the partial and total astrophysical $S$
factors for $^{6}$Li$(d,p)^{7}$Li and $^{6}$Li$(d,n)^{7}$Be, respectively.
In the $^{6}$Li$(d,p_{0})^{7}$Li channel (left panel of
Fig.~\ref{Fig:SFactors6LiDP7LiLS}), the $0^{+}$ and $1^{+}$ states dominate
for $0\leq E_{\rm cm}\lesssim 0.5$~MeV, while at higher energies the $2^{-}$
state provides the largest contribution. For $^{6}$Li$(d,p_{1})^{7}$Li (right
panel), the $1^{+}$ and $0^{+}$ components dominate below
$E_{\rm cm}\lesssim 0.25$~MeV, whereas at higher energies the $1^{-}$ and
$2^{-}$ states become dominant.

The pattern for $^{6}$Li$(d,n_{0})^{7}$Be (left panel of
Fig.~\ref{Fig:SFactors6LiDN7BeLS}) is similar: the $0^{+}$ and $1^{+}$ states
dominate at low energies ($0\leq E_{\rm cm}\lesssim 0.4$~MeV), while at
higher energies the negative--parity states become more important. The
partial $S$ factors for $^{6}$Li$(d,n_{1})^{7}$Be (right panel) closely
resemble those for $^{6}$Li$(d,p_{1})^{7}$Li, indicating that the same set of
$J^{\pi}$ states governs the population of the first excited $1/2^{-}$ level
in $^{7}$Li and $^{7}$Be. 

\subsection{Coulomb effects in the charged entrance channels
$p+{}^{7}$Li and $d+{}^{6}$Li}

Having analyzed the $S$ factors for the $^{7}$Li$(p,\alpha)^{4}$He and $^6$Li(d,$\alpha)^{4}$He
reactions separately, we now compare the impact of the Coulomb barrier in
these charged entrance channels. This comparison quantifies how the different
charges and reduced masses suppress the low--energy $S$ factors relative to
each other and helps disentangle Coulomb effects from structural
enhancements associated with the $^{8}$Be spectrum.

Coulomb repulsion plays a central role in suppressing nuclear reactions
involving charged particles at low center--of--mass energies. The strength of
this suppression is characterized by the Sommerfeld parameter
\[
\eta = \frac{Z_1 Z_2 e^2}{\hbar} \sqrt{ \frac{\mu}{2E} },
\]
where $Z_1$ and $Z_2$ are the charges of the interacting clusters, $E$ is the
center--of--mass energy, and $\mu$ is the reduced mass,
\[
\mu = m \frac{A_1 A_2}{A_1 + A_2},
\]
with $A_1$ and $A_2$ the mass numbers and $m$ the nucleon mass. For the
$p+{}^{7}$Li and $d+{}^{6}$Li systems one finds
\[
\eta_{d} = \sqrt{\frac{12}{7}}\,\eta_{p},
\]
so that the Sommerfeld parameter for $d+{}^{6}$Li is only slightly larger
than for $p+{}^{7}$Li.

Despite this relatively small difference in $\eta$, the astrophysical
$S$ factors for the $^{7}$Li$(p,\alpha)^{4}$He and $^6$Li(d,$\alpha)^{4}$He reactions differ by more than two orders of
magnitude. Experimentally, the $S$ factor for $^6$Li(d,$\alpha)^{4}$He lies in the range
16.9--23~MeV\,b at zero energy, whereas that for $^{7}$Li$(p,\alpha)^{4}$He is
$S(0)\approx 0.06$~MeV\,b, about 300 times smaller (see
Tables~\ref{Tab:SfactExpScr1} and \ref{Tab:SfactExpScr2} for polynomial
fits). This discrepancy cannot be attributed solely to the exponential
Coulomb suppression encoded in the Gamow factor. While the $S$--factor formalism removes the leading tunneling probability, it
retains residual Coulomb effects through the interference of Coulomb and
nuclear interactions, and the remaining difference reflects primarily
nuclear--structure effects: the details of the cluster--cluster interaction,
the entrance–exit channel couplings, and the pattern of subthreshold and
near--threshold $^{8}$Be resonances.

Our microscopic calculations reproduce the $S$ factor for $p+{}^{7}$Li with
good accuracy, yielding $S(0)\simeq 0.07$~MeV\,b, consistent with
experiment. For the $d+{}^{6}$Li reaction, the present model predicts
$S(0)\simeq 2.16$~MeV\,b, roughly 30~times larger than the $p+{}^{7}$Li
value but still an order of magnitude below the experimental $S(0)$ for
$d+{}^{6}$Li. Thus the model captures the main Coulomb effects and part of the structural
enhancement responsible for the much larger low--energy $S$ factor of
$d+{}^{6}$Li compared to $p+{}^{7}$Li, but it still underestimates the
experimental $d+{}^{6}$Li $S$ factor.  This remaining deficit is consistent with the
shifted $^{6}$Li+$d$ threshold and with the absence, in the present
implementation, of a broad $2^{+}$ resonance near that threshold.

\subsection{Low-energy approximations to experimental astrophysical S-factor data}

To assess the behavior of astrophysical S-factors predicted by our model at low energies, we compare them with a selection of empirical parameterizations that have been widely used in the literature. These low-energy fits are derived from experimental data, typically over sub-MeV energy ranges relevant to astrophysical processes. Although differing in functional form and fitted datasets, most aim to capture the smooth, non-resonant component of the S-factor near threshold. They are often employed in nuclear reaction rate calculations due to their simplicity and computational convenience.

However, these parameterizations are purely empirical and lack a direct theoretical foundation; their extrapolation behavior outside the data range is not always controlled. As such, while comparisons with them offer useful insight into how our model aligns with the established data landscape, they must be interpreted with appropriate caution. Our many-channel microscopic cluster model provides a physically motivated description of the same reactions, and its agreement or deviation from these fits may help identify both the strengths and the limitations of empirical approaches.

\begin{figure}[ptbh]
\begin{centering}
\includegraphics[height=10.6492cm,width=13.7618cm]{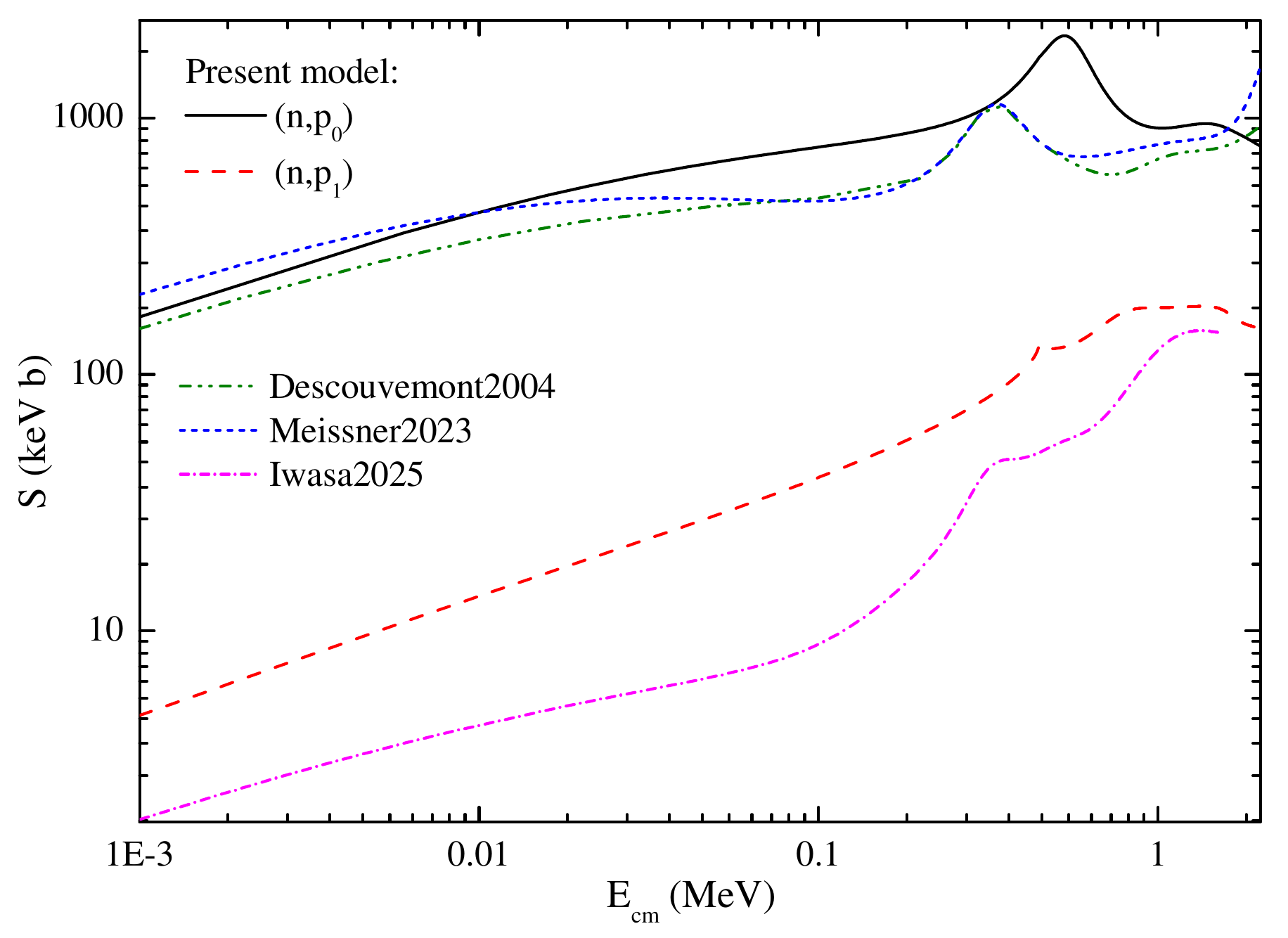}
\caption{Low-energy behavior of the astrophysical S factor for the reaction $^{7}$Be$(n,p)^7$Li, calculated within the present model and compared with analytical approximations based on experimental data fits from Refs.~\cite{2004ADNDT..88..203D,2023EPJA...59..223M,Iwasa2025}.}
\label{FIG:Sfactorn7BeAAS0}
\end{centering}
\end{figure}

\begin{figure}[ptbh]
\begin{centering}
\includegraphics[height=10.6492cm,width=13.7618cm]{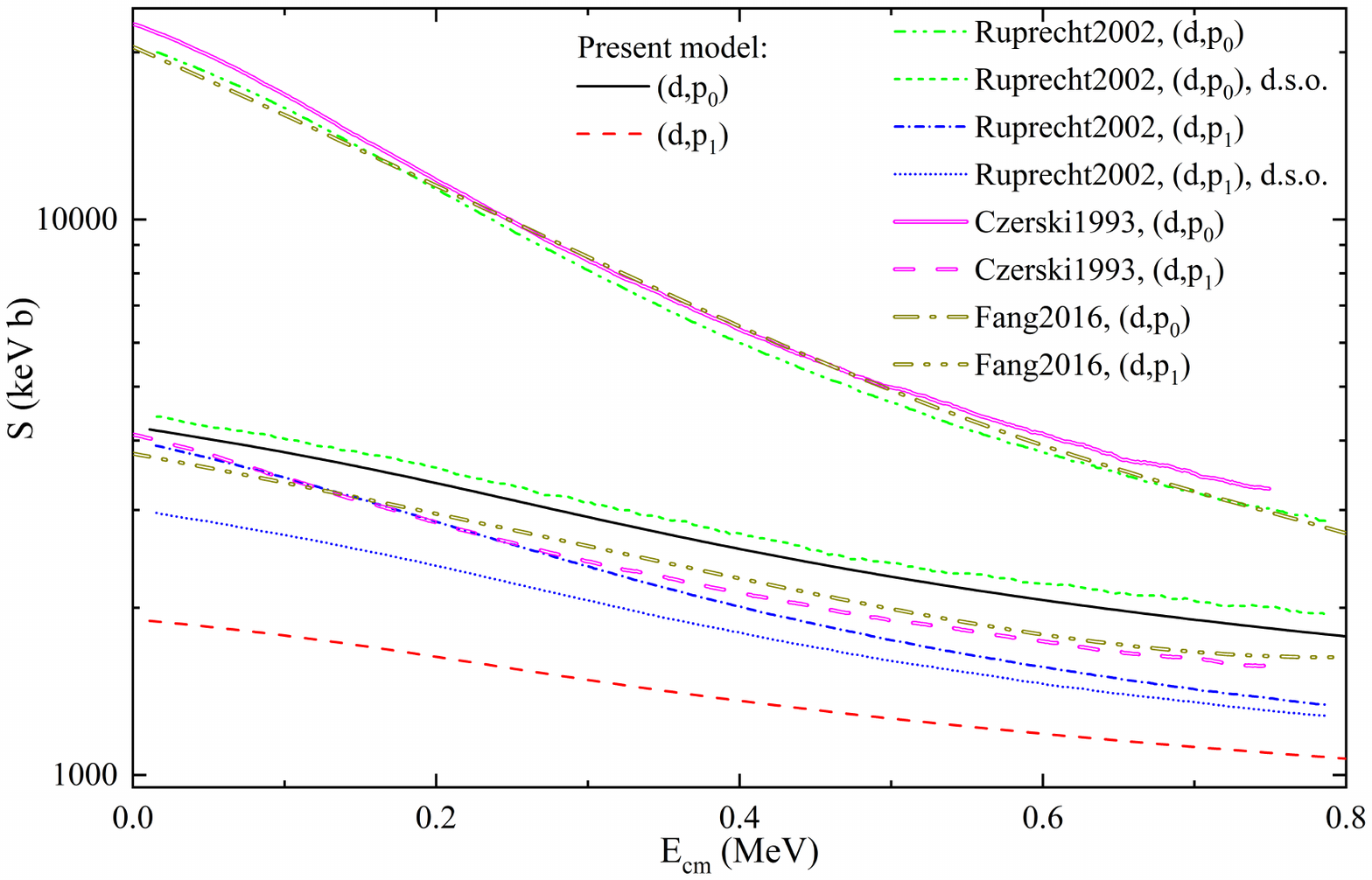}
\caption{Low-energy behavior of the astrophysical S factor for the reaction $^{6}$Li$(d,p)^7$Li, shown separately for the \( (d,p_0) \) and \( (d,p_1) \) channels. Solid and dashed curves represent the present model, while colored lines show analytical approximations based on experimental data from Refs.~\cite{RuprechtDiss,1993PhLB..307...20C,2016PhRvC..94e4602F}. For comparison, DWBA calculations including only direct contributions are also shown (“Ruprecht2002, d.s.o.”).}
\label{FIG:Sfactord6LiAAS0}
\end{centering}
\end{figure}

Tables \ref{Tab:SfactExpScr1} and \ref{Tab:SfactExpScr2} summarize several representative parameterizations of the experimental S-factor data  for the reactions $^7\mathrm{Li}(p,\alpha)^4$He and $^6\mathrm{Li}(d,\alpha)^4$He and include, for comparison, the corresponding low-energy expansions obtained from our model. We provide explicit tables only for these two channels because the literature contains a
large number of competing polynomial fits for them, so collecting the most widely used parameterizations in one place is particularly useful; for the other reactions, where only a few reference fits exist, graphical comparisons in Figs.~\ref{FIG:Sfactorn7BeAAS0}–\ref{FIG:Sfactord6LiAAS0} are sufficient. The lines labeled “Our model” in Tables~\ref{Tab:SfactExpScr1} and \ref{Tab:SfactExpScr2} represent polynomial fits to the $S(E)$ values generated by our microscopic calculation, valid over the energy ranges indicated in the table captions.
\begin{table}[tbph] \centering
\caption{The low-energy approximations to the experimental data of the astrophysical S-factors of the reaction $^7$Li(p,$\alpha$)$\alpha$. Our model fit applies to the range 6–250~keV.}
\label{Tab:SfactExpScr1}
\begin{tabular}{|c|c|c|}
\hline
Our notation& Source & S(E), MeV b \\
\hline
Engstler1992&\cite{1992ZPhyA.342..471E}&$0.0593 + 0.1929E - 0.3555E^2 + 0.2363E^3$ \\
\hline
Smith1993&\cite{Smith1993}& $0.052 + 0.041\left[ {1 - \exp \left( { - 8.804E} \right)} \right]$ \\
\hline
Yamashita1995&\cite{1995NuPhA.582..270Y}& $0.0565 + \mbox{0.1705}E - \mbox{0.1986}E^2$ \\
\hline
Lattuada2001&\cite{2001ApJ...562.1076L}& $0.055 + \mbox{0.210}E - \mbox{0.310}E^2$ \\
\hline
Barker2002&\cite{2002NuPhA.707..277B}& $0.0621 + 0.159E - 0.280E^2 + 0.186E^3$ \\
\hline
Cyburt2004&\cite{Cyburt2004}& $0.0607 + 0.1926E - 0.4603E^2 + 0.5181E^3 - 0.1951E^4$ \\
\hline
Serpico2004&\cite{Serpico2004}& $0.0609 + 0.173E - 0.319E^2 + 0.217E^3$ \\
\hline
CruzDiss&\cite{CruzDiss}& $0.0594 + 0.141E - 0.223E^2 + 0.153E^3$ \\
\hline
Kimura2007&\cite{Kimura20072a}& $0.062 + 0.15E - 0.24E^2 + 0.14E^3$ \\
\hline
Wang2012&\cite{2012JPhG...39a5201W}& $0.0616 + 0.162E - 0.284E^2 + 0.187E^3$ \\
\hline
Our model&This article& $0.0721 - 0.2143E + 2.2294E^2 - 3.9083E^3 + 4.5481E^4$ \\
\hline
\end{tabular}
\end{table}
\begin{table}[tbph] \centering
\caption{The low-energy approximations to the experimental data of the astrophysical S-factors of the reaction $^6$Li(d,$\alpha$)$\alpha$. Our model fit applies to the range 1–250~keV.}%
\label{Tab:SfactExpScr2}
\begin{tabular}{|c|c|c|}
\hline
Our notation&Source& S(E), MeV b\\
\hline
Engstler1992&\cite{1992ZPhyA.342..471E}& $\mbox{18.79} - \mbox{58.54}E + \mbox{66.64}E^2 - \mbox{25.81}E^3$ \\
\hline
Czerski1997&\cite{1997PhRvC..55.1517C}& $23\exp \left\{ { - 4.838E + 1.3586E^2} \right\}$ \\
\hline
Musumarra2001&\cite{2001PhRvC..64f8801M}& $\mbox{16.9} - \mbox{41.6}E + \mbox{28.2}E^2$ \\
\hline
Barker2002&\cite{2002NuPhA.707..277B}& $\mbox{19.7} - \mbox{66.0}E + \mbox{79.7}E^2 - \mbox{33.0}E^3$ \\
\hline
Wang2012&\cite{2012JPhG...39a5201W}& $\mbox{20.5} - \mbox{70.6}E + \mbox{88.5}E^2 - \mbox{37.9}E^3$ \\
\hline
Fang2016&\cite{2016PhRvC..94e4602F}& $\mbox{19.20} - \mbox{62.24}E + \mbox{73.13}E^2 - \mbox{29.51}E^3$ \\
\hline
Our model&This article& $2.16 - 2.87E + 0.53E^2 - 3.09E^3$ \\
\hline
\end{tabular}
\end{table}

\begin{figure}[ptbh]
\begin{centering}
\includegraphics[height=10.6492cm,width=13.7618cm]{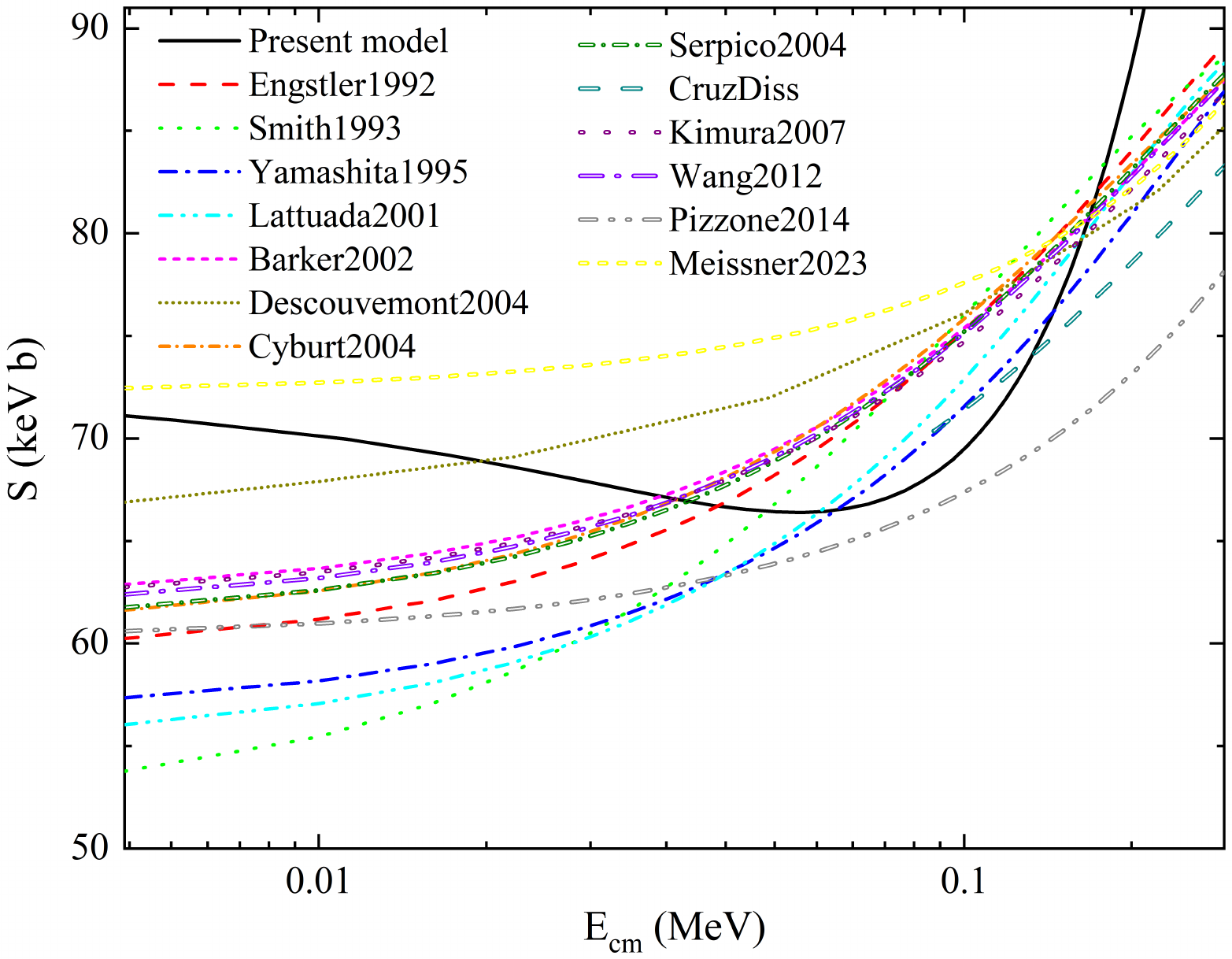}
\caption{Low-energy behavior of the astrophysical S factor for the reaction $^{7}$Li$(p,\alpha)^4$He, calculated within the present model and compared with various low-energy analytical approximations derived from experimental data and listed in Table \ref{Tab:SfactExpScr1}}
\label{FIG:SfactorP7LiAAS0}
\end{centering}
\end{figure}

Figure \ref{FIG:SfactorP7LiAAS0} shows the detailed low-energy behavior of the astrophysical S factor for the reaction $^{7}$Li$(p,\alpha)^4$He in the energy range $E \leq 250$ keV. The total S factor obtained within the present model exhibits a slow increase as the energy decreases below $E \lesssim 50$ keV. This trend originates from the subthreshold $2^{+}$ resonance located about 0.2 MeV below the $^{7}$Li + $p$ threshold, which induces a parabolic shape of the S factor in this region. The predicted behavior is consistent with low-energy analytical approximations to the experimental data shown in the figure.

Figure~\ref{FIG:Sfactorn7BeAAS0} shows a comparison of the low-energy behavior of the astrophysical S-factor for the reaction $^7$Be$(n,p)^7$Li, as calculated in the present microscopic model, with several representative results from the literature. The curve by Descouvemont et al.~\cite{2004ADNDT..88..203D} is based on an $R$-matrix analysis of low-energy experimental data, while the parametrization by Meissner \textit{et al.}~\cite{2023EPJA...59..223M} results from a fit with a nonrelativistic Breit–Wigner function and polynomials in $\sqrt{E}$ to the experimental measurements of Refs.~\cite{1988PhRvC..37..917K,1989CzJPh..39.1263C}. The fit labeled “Iwasa2025” corresponds to an S-factor extracted from the total cross section of the $^7$Be(n,p$_1$)$^7$Li reaction, as reported in~\cite{Iwasa2025}.
The present model reproduces both the absolute value and the energy dependence of the S-factor for the $^7$Be(n,p$_0$)$^7$Li reaction very well in the low-energy region ($E < 0.02$~MeV), while the predicted S-factor for the $^7$Be(n,p$_1$)$^7$Li channel exceeds the Iwasa2025 fit at the energies below 1 MeV.

Figure~\ref{FIG:Sfactord6LiAAS0} shows the low-energy astrophysical  S factors for the reactions  $^6\mathrm{Li}(d,p_0)^7\mathrm{Li}$ and  $^6\mathrm{Li}(d,p_1)^7\mathrm{Li}$, as calculated in the present model, compared to low-energy analytical fits to experimental data from Refs.~\cite{RuprechtDiss,1993PhLB..307...20C,2016PhRvC..94e4602F}. The calculated  S -factor for the \( (d,p_1) \) channel is in reasonable agreement with empirical approximations across the energy range shown, while for the \( (d,p_0) \) channel the model underestimates the low-energy enhancement, consistent with the overall underestimation of the $^{6}\mathrm{Li}(d,\alpha)^{4}\mathrm{He}$ $S$ factor discussed above. This enhancement is commonly attributed to the influence of a broad subthreshold \( 2^+ \) resonance in \( ^8\mathrm{Be} \), located approximately 80~keV below the reaction threshold~\cite{1993PhLB..307...20C}. Since this resonance is not included in our current model, the resulting  S factor for the \( (d,p_0) \) channel aligns more closely with DWBA calculations containing only the direct component~\cite{RuprechtDiss} (shown as “Ruprecht2002, \( (d,p_0) \), d.s.o.” in Fig.~\ref{FIG:Sfactord6LiAAS0}).

\subsection{Hierarchy of reaction channels in the $^{8}$Be compound system at Gamow energies}

\begin{figure}[ptbh]
\begin{center}
\includegraphics[width=\textwidth]{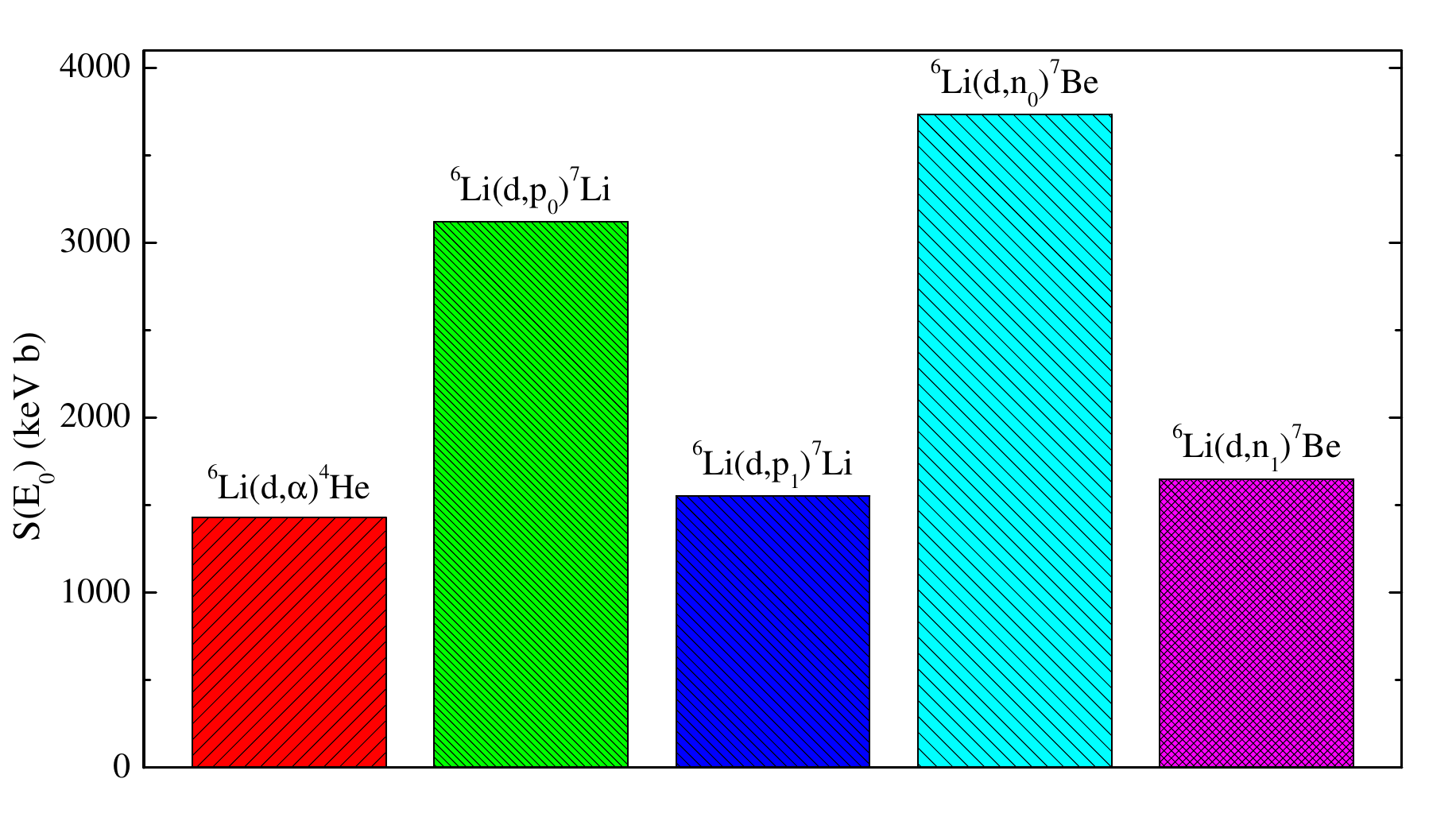}%
\caption{Astrophysical S factors for reactions induced by deuterons on $^{6}$Li, evaluated at the Gamow energy $E_{0}=250$ keV. Here $p_{0}$ ($n_{0}$) and $p_{1}$ ($n_{1}$) denote population of the ground ($3/2^{-}$) and first excited ($1/2^{-}$) states in $^{7}$Li ($^{7}$Be), respectively.}
\label{Fig:SFactors6LiD_E0}%
\end{center}
\end{figure}

Figure \ref{Fig:SFactors6LiD_E0} summarizes the hierarchy of deuteron-induced reactions on $^{6}$Li at a stellar temperature $T=0.8$ GK. The astrophysical S factors are evaluated at the corresponding Gamow energy, $E_{0}=250$ keV (see Table \ref{Tab:GamowWinds}). Within the present model, the $^{6}$Li$(d,n_{0})^{7}$Be channel is dominant: its $S(E_{0})$ exceeds those of $^{6}$Li$(d,\alpha)^4$He, $^{6}$Li$(d,p_{1})^{7}$Li, and $^{6}$Li$(d,n_{1})^{7}$Be by more than a factor of two. The second most important reaction is $^{6}$Li$(d,p_{0})^{7}$Li, whose astrophysical S factor at the Gamow energy is about 83\% of that of the dominant $^{6}$Li$(d,n_{0})^{7}$Be reaction. 
 
\begin{figure}[ptbh]
\begin{center}
\includegraphics[width=\textwidth]{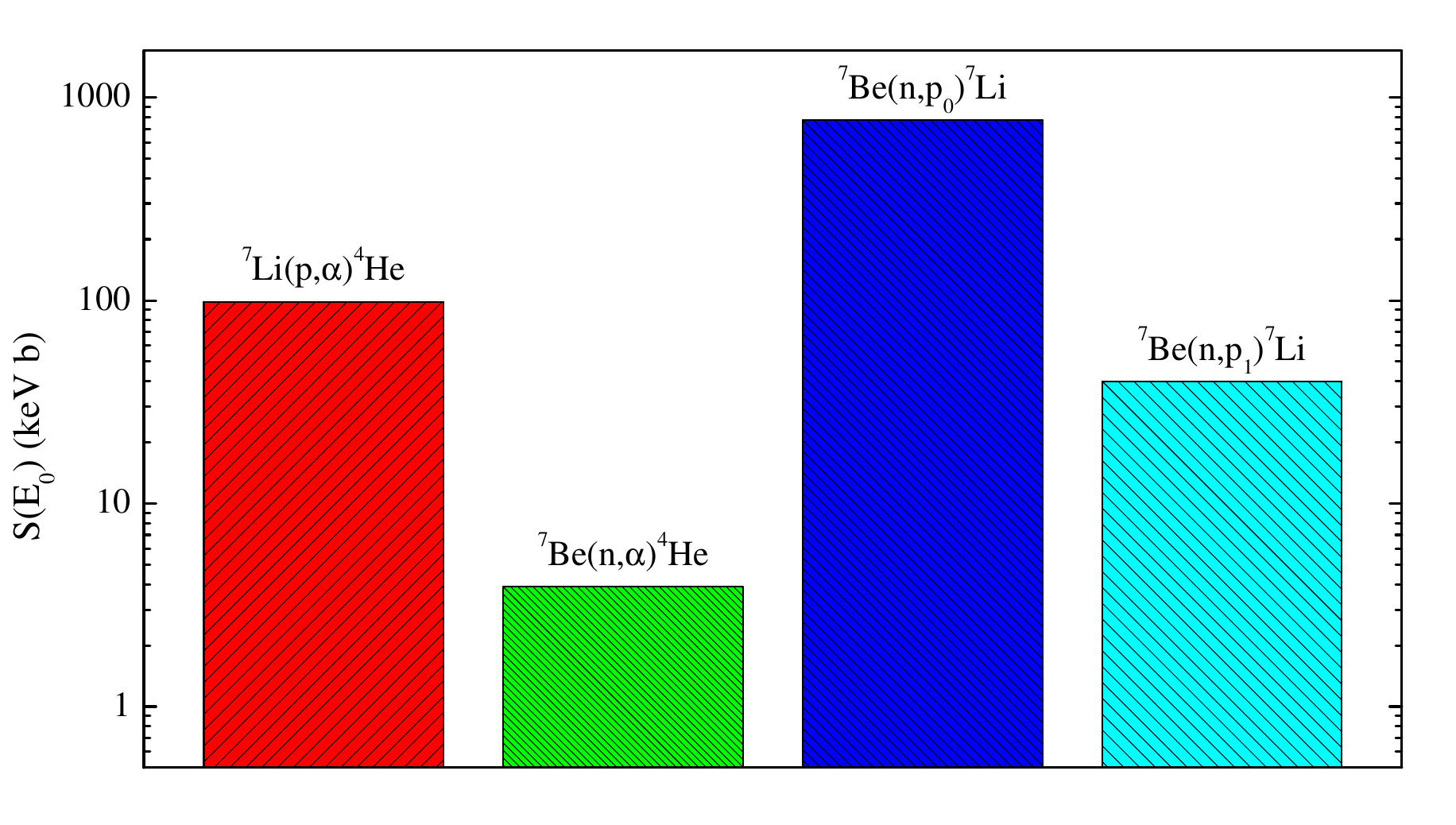}%
\caption{Astrophysical S factors for reactions induced by neutrons on $^{7}$Be, evaluated at the Gamow energy $E_{0}=103$ keV, and for the $^{7}$Li$(p,\alpha)^{4}$He reaction, evaluated at $E_{0}=209$ keV.}
\label{Fig:SFactors7BeN_E0}%
\end{center}
\end{figure}

The hierarchy of reactions induced by neutrons on $^{7}$Be, evaluated at the effective energy $E_{0}=103$~keV, is shown in Fig.~\ref{Fig:SFactors7BeN_E0}.
 The S factor of the charge-exchange reaction $^{7}$Be$(n,p_{0})^{7}$Li is about one order of magnitude larger than that of $^{7}$Be$(n,p_{1})^{7}$Li (771 keV b compared with 40 keV b) and two orders of magnitude larger than that of $^{7}$Be$(n,\alpha)^{4}$He ($\approx$ 4 keV b). Thus, at a stellar temperature of $T=0.8$~GK neutron
interactions with $^{7}$Be predominantly proceed through the $^{7}$Be$(n,p_{0})^{7}$Li channel rather than through $\alpha$ emission. For comparison, Fig. \ref{Fig:SFactors7BeN_E0} also includes the S factor of the $^{7}$Li$(p,\alpha)^{4}$He reaction, evaluated at its Gamow energy $E_{0}=209$~keV,  which is substantially smaller—by roughly a factor of eight—than that of the dominant $^{7}$Be$(n,p_{0})^{7}$Li channel.

\section{Conclusions}
\label{concl}

In this second paper of our series on the $^{8}$Be system, we extend the
microscopic many--channel three--cluster framework used in Paper~I \cite{2025PhRvC.112a4328Z} to
low--energy reactions in the $p+{}^{7}$Li, $n+{}^{7}$Be, and $d+{}^{6}$Li
entrance channels. Within this unified model,
which includes the $^{4}$He+$^{3}$H+$p$, $^{4}$He+$^{3}$He+$n$,
$^{4}$He+$d$+$d$, and $^{4}$He+$2p$+$2n$ configurations and embeds all
binary rearrangement channels with ground and excited states of $^{7}$Be,
$^{7}$Li, $^{5}$Li, and $^{5}$He, we have
calculated the astrophysical $S$ factors for the reactions
$^{7}$Li$(p,\alpha)^{4}$He, $^{7}$Be$(n,\alpha)^{4}$He,
$^{7}$Be$(n,p)^{7}$Li, $^{6}$Li$(d,\alpha)^{4}$He, $^{6}$Li$(d,p)^{7}$Li, and
$^{6}$Li$(d,n)^{7}$Be in the energy range relevant for primordial and stellar
nucleosynthesis. A dedicated scan of the low–energy region in all three
entrance channels revealed no additional narrow or broad resonances close to
the $p+{}^{7}$Li, $n+{}^{7}$Be, or $d+{}^{6}$Li thresholds; within the present
framework the often–invoked scenario of an unknown near–threshold resonance
that could resolve the lithium problems is therefore not supported. 
This conclusion is consistent with the BBN response analysis of Ref.~\cite{Broggini2012}, which finds that the enhancement of the
$^{7}$Be+n destruction rate required to solve the lithium problem is incompatible with general nuclear–physics constraints on the relevant cross sections.

For the mirror reactions \mbox{$^{7}$Li(p,$\alpha$)$^{4}$He} and \mbox{$^{7}$Be(n,$\alpha$)$^{4}$He},
and for the charge--exchange channel \mbox{$^{7}$Be(n,p)$^{7}$Li}, the calculated $S(E)$ curves reproduce both the absolute scale and the low--energy trends of the data within their quoted uncertainties. The
partial--wave analysis connects these observables directly to the
$^{8}$Be spectrum obtained in Paper~I: the low--energy
\mbox{$^{7}$Li(p,$\alpha$)$^{4}$He} $S$ factor is governed by the
$J^{\pi}=0^{+}$ component with curvature induced by subthreshold $2^{+}$
strength, whereas \mbox{$^{7}$Be(n,$\alpha$)$^{4}$He} is dominated by the $2^{+}$ wave, with the $0^{+}$ component remaining subdominant and the $4^{+}$ contribution becoming noticeable only at higher energies.
In the \mbox{$^{7}$Be(n,$\alpha$)$^{4}$He} channel, reactions on the first
excited $1/2^{-}$ state of $^{7}$Be contribute more strongly than those on
the ground $3/2^{-}$ state, in line with the state--resolved cross sections
of Ref.~\cite{2017PhRvL.118e2701K}. For $^{7}$Be$(n,p)^{7}$Li the enhancement
near $E_{\mathrm{cm}}\sim 0.5$~MeV is generated by the $3^{+}$ and $1^{-}$
resonances identified previously, and cluster polarization is essential to
reproduce both the low–energy normalization and the correct position of the
$(n,p_{0})$ peak.  Thus, the same microscopic ingredients --- specific $J^{\pi}$ resonances and
polarization of the binary subsystems --- that were shown in
Paper~I to shape the high--lying $^{8}$Be spectrum
also control the astrophysical $S$ factors at low entrance--channel
energies near the $p+{}^{7}$Li, $n+{}^{7}$Be, and $d+{}^{6}$Li thresholds.

For deuteron–induced reactions on $^{6}$Li the model captures the qualitative hierarchy of exit channels and the dominant $J^{\pi}$ components but underestimates the absolute $S$ factors at $E_{\mathrm{cm}}\lesssim 1$~MeV. This shortcoming can be traced to the position of the $^{6}$Li+d threshold and to the absence, in the present implementation, of a broad subthreshold $J^{\pi}=2^{+}$ structure in $^{8}$Be very close to that threshold. Cluster polarization enhances the
$^{6}$Li$(d,\alpha)^{4}$He $S$ factor and improves agreement with experiment,
while having a more modest effect on $^{6}$Li$(d,p)^{7}$Li and
$^{6}$Li$(d,n)^{7}$Be. Despite this limitation, the $J^{\pi}$ analysis remains physically transparent: $0^{+}$ (and, at higher energies, $2^{+}$) controls reactions with the $\alpha+\alpha$ exit channel, whereas
combinations of $0^{+}$, $1^{+}$ and negative–parity states ($1^{-},2^{-}$)
govern the $(d,p)$ and $(d,n)$ transitions to the ground and first excited
states of $^{7}$Li and $^{7}$Be. Comparisons with empirical low–energy fits
confirm the microscopic description for $^{7}$Li$(p,\alpha)^{4}$He and
$^{7}$Be$(n,p_{0})^{7}$Li and isolate the missing subthreshold $2^{+}$ strength
as the main source of discrepancy in $^{6}$Li$(d,\alpha)^{4}$He and
$^{6}$Li$(d,p_{0})^{7}$Li, while the $^{6}$Li$(d,p_{1})^{7}$Li channel, where
this strength is known to be much weaker, is reproduced more accurately.

Evaluating the $S$ factors at Gamow energies characteristic of Big Bang and
stellar conditions, we find that $^{6}$Li$(d,n_{0})^{7}$Be and
$^{6}$Li$(d,p_{0})^{7}$Li dominate among the deuteron–induced channels on
$^{6}$Li at $T=0.8$~GK, while $^{7}$Be$(n,p_{0})^{7}$Li is by far the strongest
neutron–induced channel on $^{7}$Be and exceeds the
$^{7}$Li$(p,\alpha)^{4}$He $S$ factor evaluated at its Gamow energy. These hierarchies show that, at the relevant Gamow energies, neutron--induced reactions on $^{7}$Be and deuteron--induced reactions on $^{6}$Li provide
significant pathways for the production and redistribution of $^{7}$Li and $^{7}$Be.
 Taken together with the resonance analysis of Paper~I,
the present results demonstrate that a single microscopic, many--channel cluster
approach can give a coherent and predictive description of both the $^{8}$Be
spectrum and reaction observables at low entrance--channel energies, and they
highlight specific observables --- such as the relative contributions of the ground and first excited states in
$^{7}$Be(n,$\alpha$)$^{4}$He and $^{7}$Be(n,p)$^{7}$Li, as well as the
hierarchy of $S(E_{0})$ values at Gamow energies --- as concrete targets for
future high--precision measurements of lithium--related reaction channels.
\begin{acknowledgments}

This work received partial support from the Program of Fundamental Research of the Physics and Astronomy Department of the National Academy of Sciences of Ukraine (Project No. 0122U000889). We extend our gratitude to the Simons Foundation for their financial support (Award ID: SFI-PD-Ukraine-00014580).   
\end{acknowledgments}

\appendix

\section{Experimental results \label{Sec:Exper}}

In this Appendix we compile the experimental data used for comparison with our theoretical results. 
The datasets are summarized in six tables: 
Tables~\ref{Tab:SfactExpP7LiAA}, \ref{Tab:Exp7BeNAA}, \ref{Tab:Exp7BeNP7Li},
\ref{Tab:SfactExp62AA}, \ref{Tab:SfactExp62P7Li}, and \ref{Tab:SfactExp62N7Be},
which correspond to the reactions 
$^{7}$Li$(p,\alpha){}^{4}$He, 
$^{7}$Be$(n,\alpha){}^{4}$He, 
$^{7}$Be$(n,p){}^{7}$Li, 
$^{6}$Li$(d,\alpha){}^{4}$He, 
$^{6}$Li$(d,p){}^{7}$Li, and $^{6}$Li$(d,n){}^{7}$Be, respectively. 

Each table lists the measured (or derived) quantity, its energy range, and the corresponding reference.  
The column “EXFOR source’’ indicates whether the numerical data were taken directly from the EXFOR database (“+’’) or digitized from the original publication (“--’’).  
Dataset labels given under “Our notation’’ are identical to those used in the figures.  
For recent experiments not yet included in EXFOR, we follow the same referencing format.

The conventions used in compiling the data and the specific features of individual experiments are outlined below.

\subsection*{Notes on experimental datasets and conventions}

\noindent\textbf{General conventions.} All energies are quoted in the center-of-mass frame \(E_{\rm cm}\). Reported quantities are the integrated cross section \(\sigma(E)\), differential cross section \(d\sigma/d\Omega\), or the astrophysical \(S\) factor \(S(E)\) in the units given in the tables. When the literature reports \(S\) in MeV\,b, we convert to keV\,b for internal uniformity in plots; the tables preserve the authors’ original units. The “EXFOR source” column indicates whether numerical data were taken directly from EXFOR (“+”) or from the article/thesis tables or our digitization (“--”). The “Our notation” column lists the short labels used in our figures.

\noindent\textbf{Electron screening and indirect methods.} At very low energies in charged-particle reactions, some direct measurements are affected by electron screening; we compare to the values as reported by the authors without applying additional screening corrections. Indirect determinations (e.g., Trojan-Horse or breakup methods) are largely free from atomic screening but may carry model dependence.

\noindent\textbf{Practical implications for our comparisons.} 
(i) No cross-experiment renormalization has been applied in the Appendix tables; each dataset is shown as published (units and targets preserved). 
(ii) Forward–reverse conversions (for \(^{7}\)Be(n,p)) follow the specific relations used by the cited authors; we do not recompute these from raw reverse-channel data here.

\begin{table}[tbph] \centering
\caption{Experimental data for the reaction $^7$Li(p,$\alpha$)$^4$He.}%
\begin{tabular}{|c|c|c|c|c|c|}
\hline Reaction & Quantity & $E_{cm}$, MeV & EXFOR source & Our notation & Ref. \\
\hline (p,$\alpha )$ & $\sigma \left( E\right) $, mb & 0.020-0.044
& + & Fiedler1967 & \cite{Fiedler1967} \\
\hline (p,$\alpha )$ & $\sigma \left( E\right) $, mb & 0.044-0.105
& + & Lee1969 & \cite{Lee1969} \\
\hline (p,$\alpha )$ & $\sigma \left( E\right) $, mb & 0.114-0.490
& + & Spinka1971 & \cite{Spinka1971, Spinka1971er} \\
\hline (p,$\alpha )$ & $S\left(E\right) $, keV b & 0.025-0.873
& -- & Rolfs1986 & \cite{1986NuPhA.455..179R} \\
\hline (p,$\alpha )$ & $S\left( E\right) $, MeV b & 0.013-0.041
& + & Engstler1989 & \cite{Engstler1989} \\
\hline (p,$\alpha )$ & $\sigma \left(E\right) $, b & 0.017-0.219
& + & Harmon1989 & \cite{Harmon1989} \\
\hline (p,$\alpha )$ & $S\left( E\right) $, keV b & 0.010-0.065
& + & Schroder1989 & \cite{Schroder1989} \\
\hline (p,$\alpha )$ & $S\left( E\right) $, keV b & 0.013-1.004
& + & Engstler1992 & \cite{1992ZPhyA.342..471E} \\
\hline (p,$\alpha )$ & $S\left( E\right) $, keV b & 0.010-0.290
& + & Spitaleri1999 & \cite{1999PhRvC..60e5802S} \\
\hline (p,$\alpha )$ & $S\left( E\right) $, mb & 0.035-0.087
& + & Spraker1999 & \cite{Spraker1999} \\
\hline (p,$\alpha )$ & $S\left( E\right) $ , keV b & 0.009-0.290
& -- & Pellegriti2000 & \cite{Pellegriti2000} \\
\hline (p,$\alpha )$ & $S\left( E\right) $, keV b & 0.010-0.371
& + & Lattuada2001 & \cite{2001ApJ...562.1076L} \\
\hline (p,$\alpha )$ & $S\left( E\right) $, keV b & 0.025-0.083
& + & Cruz2005 & \cite{2005PhLB..624..181L, CruzDiss} \\
\hline (p,$\alpha )$ & $S\left( E\right) $, keV b & 0.090-1.740
& + & Cruz2009 & \cite{2009NIMPB.267..478C, CruzDiss} \\
\hline (p,$\alpha )$ & $S\left(E\right) $, MeV b & 0.017-0.053
& + & Fang2011 & \cite{2011JPSJ...80h4201F} \\
\hline (p,$\alpha )$ & $S\left(E\right) $, MeV b & 0.087-0.207
& -- & Chen2014 & \cite{Chen2014} \\
\hline (p,$\alpha )$ & S(E), keV b & 0.030-0.258
& -- & Vesic2014 & \cite{Vesic2014}\\
\hline
\end{tabular}
\label{Tab:SfactExpP7LiAA}%
\end{table}
\vspace{-0.4em} 
\noindent\textbf{Comments on $^{7}$Li(p,$\alpha$)$^{4}$He data}

Spinka (1971) and Harmon (1989) report the number of \(\alpha\) particles emitted per incident proton. Since \(p+{}^{7}\mathrm{Li}\to 2\alpha\) produces two \(\alpha\) particles per event, the total reaction cross section is obtained by dividing the reported \(\alpha\) yield by 2 before converting to \(\sigma(E)\).\\
\begin{table}[tbph] \centering
\caption{Experimental cross section of the reaction $^7$Be(n,$\alpha$)$^4$He.}%
\begin{tabular}{|c|c|c|c|c|c|}
\hline Reaction & Quantity & $E_{cm}$, MeV & EXFOR source & Our notation & Ref. \\
\hline $^{7}$Be(n,$\alpha $)$^{4}$He & $\sigma (E)$, mb & 0.011-5.754
& -- & Hou2015 & \cite{2015PhRvC..91e5802H} \\
\hline $^{7}$Be$^{\ast }$(n,$\alpha $)$^{4}$He & $\sigma (E)$, mb & 0.204-0.376
& + & Kawabata2017* & \cite{2017PhRvL.118e2701K} \\
\hline $^{7}$Be$_{gs}$(n,$\alpha $)$^{4}$He & $\sigma (E)$, mb & 0.235-0.805 
& + & Kawabata2017gs & \cite{2017PhRvL.118e2701K} \\
\hline $^{7}$Be(n,$\alpha $)$^{4}$He & $\sigma (E)$, mb & 0.087-4.048
& -- & Lamia2017, 2H & \cite{2017ApJ...850..175L} \\
\hline $^{7}$Be(n,$\alpha $)$^{4}$He & $\sigma (E)$, mb & 0.104-5.252
& -- & Lamia2017, 3He & \cite{2017ApJ...850..175L} \\
\hline $^{7}$Be(n,$\alpha $)$^{4}$He & $\sigma (E)$, mb & 0.027-1.700
& + & Lamia2019 & \cite{2019ApJ...879...23L} \\
\hline $^{7}$Be(n,$\alpha $)$^{4}$He & $\sigma (E)$, mb & 0.045-1.955
& + & Hayakawa2021 & \cite{2021ApJ...915L..13H} \\
\hline $^{7}$Be(n,$\alpha $)$^{4}$He & $\sigma (E)$, mb & 0.029-1.918
& -- & Lagni2021 & \cite{LagniDiss} \\
\hline
\end{tabular}
\label{Tab:Exp7BeNAA}
\end{table}
\vspace{-0.4em} 

\noindent\textbf{Comments on \(^{7}\mathrm{Be}(n,\alpha){}^{4}\mathrm{He}\) data}

\emph{State selection (Kawabata 2017).} Separate cross sections are reported for reactions on ground-state \(^{7}\)Be and on the first excited state \(^{7}\)Be\(^*\) (0.429\,MeV). We list these datasets distinctly as “Kawabata2017gs” and “Kawabata2017*”.\\
\emph{Indirect extractions (Lamia 2017).} The \(^{7}\)Be(n,\(\alpha\)) cross section was derived using deuteron and \({}^{3}\)He breakup data (“2H” and “3He” rows), via detailed balance and reaction theory. \\
\begin{table}[tbph] \centering
\caption{Experimental data for the reaction $^7$Be(n,p)$^7$Li.}%
\begin{tabular}{|c|c|c|c|c|c|}
\hline Reaction & Quantity & $E_{cm}$, MeV & EXFOR source & Our notation & Ref. \\
\hline $^{7}$Be(n,p$_{0}$)$^{7}$Li & $\sigma (E)$$\times \sqrt{E}$, b MeV$^{1/2}$ & 0.0098-0.421
& -- & Gibbons1959 & \cite{DeSouza2020} \\
\hline $^{7}$Be(n,p)$^{7}$Li & $\sigma (E)$, b & 2.35$\times $10$^{-8}$-0.012
& + & Koehler1988 & \cite{1988PhRvC..37..917K} \\
\hline $^{7}$Be(n,p$_{0}$)$^{7}$Li & $\sigma (E)$$\times \sqrt{E}$, b MeV$^{1/2}$ & 0.0020-0.0094
& -- & MartinHernandez2019 & \cite{DeSouza2020} \\
\hline $^{7}$Be(n,p$_{0}$)$^{7}$Li & $\sigma (E)$, mb & 0.171-0.621
& + & Hayakawa2021, (n,p$_{0}$) & \cite{2021ApJ...915L..13H} \\
\hline $^{7}$Be(n,p$_{1}$)$^{7}$Li & $\sigma (E)$, mb & 0.05-2.45
& + & Hayakawa2021, (n,p$_{1}$) & \cite{2021ApJ...915L..13H} \\
\hline $^{7}$Be(n,p$_{0}$)$^{7}$Li & $S(E)$, keV b & 1.168-7.058
& -- & Borchers1963, (n,p$_{0}$) & \cite{Borchers1963} \\
\hline $^{7}$Be(n,p$_{1}$)$^{7}$Li & $S(E)$, keV b & 1.160-7.066
& -- & Borchers1963, (n,p$_{1}$) & \cite{Borchers1963} \\
\hline $^{7}$Be(n,p)$^{7}$Li & $S(E)$, keV b & 0.183-1.678
& -- & Burke1974 & \cite{Burke1974} \\
\hline $^{7}$Be(n,p)$^{7}$Li & $S(E)$, keV b & 0.0076-2.031
& -- & Sekharan1976 & \cite{Sekharan1976} \\
\hline $^{7}$Be(n,p)$^{7}$Li & $S(E)$, keV b & 0.147-0.977
& -- & Kumar2012 & \cite{Kumar2012} \\
\hline
\end{tabular}
\label{Tab:Exp7BeNP7Li}%
\end{table}%

\vspace{-0.4em} 
\noindent\textbf{Comments on \(^{7}\mathrm{Be}(n,p){}^{7}\mathrm{Li}\) data}

\emph{Reverse-reaction mapping (Gibbons 1959 and MartinHernandez 2019).} In Ref.  \cite{DeSouza2020}, the quantities \(\sigma(E)\sqrt{E}\) for \(^{7}\)Be(n,p\(_0\))\(^{7}\)Li were inferred from the reverse reaction \(^{7}\)Li(p,n)\(^{7}\)Be.
We adopt the tabulated values from Tables A3–A4 of Ref. \cite{DeSouza2020} for the Gibbons (1959) and Martín-Hernández (2019) inputs.\\
\emph{Legacy \(S\) factors from reverse data.} For Borchers (1963), Burke (1974), Sekharan (1976), and Kumar (2012) we list \(S(E)\) values obtained via the reverse channel using the relations summarized in \cite{DeSouza2020}.
\begin{table}[tbph] \centering
\caption{Experimental data for the reaction $^6$Li(d,$\alpha$)$^4$He.}%
\begin{tabular}{|c|c|c|c|c|c|}
\hline Reaction & Quantity & $E_{cm}$, MeV & EXFOR source & Our notation & Ref. \\
\hline (d,$\alpha )$ & $\frac{d\sigma }{d\Omega }$, mb/sr & 0.146-1.191
& + & Whaling1950 & \cite{1950PhRv...79..258W} \\
\hline (d,$\alpha )$ & $\frac{d\sigma }{d\Omega }$, $\mu $b/sr & 0.022-0.188
& + & Sawyer1953 & \cite{Sawyer1953} \\
\hline (d,$\alpha )$ & $\frac{d\sigma }{d\Omega }$, mb/sr & 0.041-0.332
& + & Hirst1954 & \cite{Hirst1954} \\
\hline (d,$\alpha )$ & $\sigma (E)$, mb & 0.674-3.708
& + & Jeronymo1962 & \cite{1962NucPh..38...11J} \\
\hline (d,$\alpha )$ & $\sigma (E)$, mb & 0.225-0.749
& + & Bertrand1968 & \cite{Bertrand1968} \\
\hline (d,$\alpha _{0})$ & $\frac{d\sigma}{d\Omega }$, mb/sr & 0.048-0.097
& + & Kato1972 & \cite{Kato1972} \\
\hline (d,$\alpha )$ & $\sigma (E)$, mb & 0.357-2.572
& + & McClenahan1975 & \cite{1975PhRvC..11..370M} \\
\hline (d,$\alpha )$ & $\sigma (E)$, mb & 0.088-0.730
& + & Elwyn1977 & \cite{1977PhRvC..16.1744E} \\
\hline (d,$\alpha )$ & $\sigma (E)$, mb & 0.075-0.135
& -- & Szabo1982 & \cite{Szabo1982} \\
\hline (d,$\alpha )$ & $\sigma (E)$, mb & 0.375-1.858
& + & Dunjiu1985 & \cite{Dunjiu1985} \\
\hline (d,$\alpha )$ & $S(E)$, MeV b & 0.037-0.952
& + & Engstler1992-1 & \cite{1992ZPhyA.342..471E} \\
\hline (d,$\alpha )$ & $S(E)$, MeV b & 0.016-0.090
& + & Engstler1992-2 & \cite{1992ZPhyA.342..471E} \\
\hline (d,$\alpha )$ & $S(E)$, MeV b & 0.014-0.073
& + & Engstler1992-3 & \cite{1992ZPhyA.342..471E} \\
\hline (d,$\alpha )$ & $S(E)$, MeV b & 0.039-0.967
& -- & Cherubini1996 & \cite{1996ApJ...457..855C} \\
\hline (d,$\alpha )$ & $S(E)$, MeV b & 0.041-0.132
& -- & Czerski1997 & \ \cite{1997PhRvC..55.1517C} \\
\hline (d,$\alpha )$ & $S(E)$, MeV b & 0.037-0.613
& -- & Pizzone2000 & \cite{2000AIPC..513..385P} \\
\hline (d,$\alpha )$ & $S(E)$, MeV b & 0.012-0.737
& + & Spitaleri2001 & \cite{2001PhRvC..63e5801S} \\
\hline (d,$\alpha )$ & $S(E)$, MeV b & 0.041-0.132
& -- & Ruprecht2004 & \cite{Ruprecht2004} \\
\hline (d,$\alpha )$ & $S(E)$, MeV b & 0.075
& + & Lalremruata2009 & \cite{2009PhRvC..80d4617L} \\
\hline (d,$\alpha )$ & $S(E)$, MeV b & 0.014-0.039
& + & Fang2011 & \cite{2011JPSJ...80h4201F} \\
\hline (d,$\alpha )$ & $\sigma (E)$, mb & 0.430-4.963
& + & Pizzone2011 & \cite{2011PhRvC..83d5801P} \\
\hline (d,$\alpha )$ & $S(E)$, MeV b & 0.023-0.052
& + & Fang2016 & \cite{2016PhRvC..94e4602F} \\
\hline (d,$\alpha )$ & $\frac{d\sigma
}{d\Omega }$, mb/sr & 1.498-7.491
& + & Paneru2024 & \cite{2024PhRvC.110d4603P} \\
\hline
\end{tabular}
\label{Tab:SfactExp62AA}%
\end{table}
\begin{table}[tbph] \centering
\caption{Experimental data for the reaction $^6$Li(d,p)$^7$Li.}%
\begin{tabular}{|c|c|c|c|c|c|}
\hline Reaction & Quantity & $E_{cm}$, MeV & EXFOR source & Our notation & Ref. \\
\hline (d,p$_{0,1})$ & $\frac{d\sigma }{d\Omega}$, mb/sr & 0.203-1.336; 0.234-1.054
& + & Whaling1950 &\cite{1950PhRv...79..258W} \\
\hline (d,p) & $\frac{d\sigma}{d\Omega }$, $\mu $b/sr & 0.022-0.188
& + & Sawyer1953 & \cite{Sawyer1953} \\
\hline (d,p$_{0,1})$ & $\sigma (E)$, mb & 0.749-1.498
& -- & Bruno1966 & \cite{1975PhRvC..11..370M} \\
\hline (d,p$_{0,1})$ & $\sigma (E)$, mb & 0.225-0.749
& + & Bertrand1968 & \cite{Bertrand1968} \\
\hline (d,p$_{0,1})$ & $\frac{d\sigma }{d\Omega}$, mb/sr & 1.522-8.316; 1.563-8.166
& + & Durr1968 & \cite{1968NuPhA.122..153D} \\
\hline (d,p$_{0,1})$ & $\sigma (E)$, mb & 1.698-5.225; 1.671-4.501
& + & Gould1975 & \cite{Gould1975} \\
\hline (d,p$_{0,1})$ & $\sigma (E)$, mb & 0.359-2.582; 0.365-2.566
& + & McClenahan1975 & \cite{1975PhRvC..11..370M} \\
\hline (d,p$_{0,1})$ & $\sigma (E)$, mb & 0.088-0.730
& + & Elwyn1977 & \cite{1977PhRvC..16.1744E} \\
\hline (d,p) & $S(E)$, MeV b & 0.036-0.161
& -- & Cecil1981 & \cite{1981PhRvC..24.1769C} \\
\hline (d,p) & $\sigma (E)$, mb & 0.075-0.135
& -- & Szabo1982 & \cite{Szabo1982} \\
\hline (d,p$_{0,1})$ & $\sigma (E)$, mb & 0.375-1.873; 0.779-1.873
& + & Dunjiu1985 & \cite{Dunjiu1985} \\
\hline (d,p$_{1})$ & $S(E)$, MeV b & 0.083-0.127
& + & Czerski1993 & \cite{1993PhLB..307...20C} \\
\hline (d,p$_{0})$ & $S(E)$, MeV b & 0.048-0.102
& + & Czerski1997 & \cite{Czerski1997} \\
\hline (d,p$_{0+1})$ & $S(E)$, MeV b & 0.0749
& + & Lalremruata2009 & \cite{2009PhRvC..80d4617L} \\
\hline (d,p$_{0,1})$ & $S(E)$, MeV b & 0.024-0.052; 0.023-0.052
& + & Fang2016 & \cite{2016PhRvC..94e4602F} \\
\hline (d,p$'\gamma )$ & $\frac{d\sigma}{d\Omega }$, mb/sr & 0.739-1.642
& + & Taimpiri2023 & \cite{2023NIMPB.539..162T} \\
\hline (d,p$_{0,1})$ & $\frac{d\sigma }{d\Omega }$, mb/sr & 1.498-7.491
& + & Paneru2024 & \cite{2024PhRvC.110d4603P}] \\ \hline
\end{tabular}
\label{Tab:SfactExp62P7Li}%
\end{table}
\vspace{-0.4em} 

\noindent\textbf{Comments on \(^{6}\mathrm{Li}(d,p){}^{7}\mathrm{Li}\) and \(^{6}\mathrm{Li}(d,n){}^{7}\mathrm{Be}\) data}

\noindent\emph{Reprinted data.} Ref. \cite{1975PhRvC..11..370M} includes the Bruno (1966) data from \cite{Bruno1966}.\\
The paper \cite{1979NSE....71..280R} reports the Barr (1975) data from \cite{Barr1975}.

\noindent\emph{Angular vs.\ total cross sections.} 
Taimpiri (2023) reported \(d\sigma/d\Omega\) at several angles. 
We used the \(0^{\circ}\) data, multiplied by \(4\pi\) to obtain the total cross section for the $S$-factor evaluation.

\begin{table}[tbp] \centering
\caption{Experimental data for the reaction $^6$Li(d,n)$^7$Be.}%
\begin{tabular}{|c|c|c|c|c|c|}
\hline Reaction & Quantity & $E_{cm}$, MeV & EXFOR source & Our notation & Ref. \\
\hline (d,n) & $\sigma (E)$, mb & 0.086-0.253
& + & Hirst1954 & \cite{Hirst1954} \\
\hline (d,n) & $\sigma (E)$, mb & 0.164-0.726
& + & Barr1975 & \cite{1979NSE....71..280R} \\
\hline (d,n$_{0,1}$) & $\sigma (E)$, mb & 0.368-2.188; 0.360-2.197
& + & McClenahan1975 & \cite{1975PhRvC..11..370M} \\
\hline (d,n$_{0,1}$) & $\sigma (E)$, mb & 0.153-0.654
& + & Elwyn1977 & \cite{1977PhRvC..16.1744E} \\
\hline (d,n) & $\sigma (E)$, mb & 0.225-0.697
& + & Ruby1979 & \cite{1979NSE....71..280R} \\
\hline (d,n$_{1}\gamma $) & $S(E)$, MeV b & 0.047-0.119
& + & Cecil1982 & \cite{Cecil1982} \\
\hline (d,n) & $\sigma (E)$, mb & 0.075-0.135
& -- & Szabo1982 & \cite{Szabo1982} \\
\hline (d,n) & $S(E)$, MeV b & 0.0966
& + & Hofstee2001 & \cite{2001NuPhA.688..527H} \\
\hline (d,n$\gamma $) & $\frac{d\sigma }{d\Omega }$, mb/sr & 0.593-1.495
& + & Aslani2023 & \cite{2023NIMPB.535...96A} \\
\hline (d,n$'\gamma$) & $\frac{d\sigma }{d\Omega }$, mb/sr & 0.739-1.642
& + & Taimpiri2023 & \cite{2023NIMPB.539..162T} \\
\hline (d,n$_{0,1}$) & $\sigma (E)$, mb & 1.498-7.491
& + & Paneru2024 & \cite{2024PhRvC.110d4603P} \\ \hline
\end{tabular}
\label{Tab:SfactExp62N7Be}%
\end{table}%

\clearpage

\bibliography{8Be}
\bibliographystyle{apsrev4-2}

\end{document}